\documentclass[12pt,letterpaper]{article}
\usepackage{amsmath,amssymb,pgf,pgfarrows,pgfnodes,float,appendix, hyperref}
\usepackage{graphicx}
\usepackage{subfigure}
\usepackage[margin=0.9in]{geometry}

\newcommand{\be}{\begin{equation}}
\newcommand{\ee}{\end{equation}}
\newcommand{\bea}{\begin{eqnarray}}
\newcommand{\eea}{\end{eqnarray}}

\title{{\rm\footnotesize \qquad \qquad \qquad \qquad \qquad \ \qquad \qquad \qquad \ \ \ \ \ \                      RUNHETC-2014-05  SCIPP 14/04}\vskip.5in     Supersymmetry Breaking and the Cosmological Constant\\ {\it To Appear in Int. J. Mod. Phys. A}}
\author{Tom Banks\\
Department of Physics and SCIPP\\
University of California, Santa Cruz, CA 95064\\
{\it and}\\
Department of Physics and NHETC\\
Rutgers University, Piscataway, NJ 08854\\
E-mail: \href{mailto:banks@scipp.ucsc.edu}{banks@scipp.ucsc.edu}
\\
}
\date{}
\begin{document}
\maketitle

\begin{abstract}
I review three attempts to explain the small value of the cosmological constant, and their connection to SUSY breaking.  They are The String Landscape, Supersymmetric Large Extra Dimensions (SLED), and the Holographic Space-time Formalism invented by Fischler and myself.
\end{abstract}

\section{Introduction}

The value of Einstein's cosmological constant (c.c.) $\Lambda$ is perhaps the most perplexing of all the naturalness or hierarchy problems in theoretical high energy physics.  Observation appears to imply\footnote{There are both observational issues, and alternative theoretical models, which prevent us from saying definitively that the observed behavior of distant supernovae and other cosmological data, give us a definite measurement of a positive value for the c.c. .   I will ignore those issues in this review.} that it is positive and has the value $10^{- 123}$ in Planck units.  The Planck length is $L_P = 10^{-33}$ cm. and the corresponding energy scale ($\hbar = c = 1$) is $M_P = 10^{19}$ GeV. The c.c. has dimensions of an energy density.  The Hubble scale associated with the c.c., which we will eventually call the de Sitter (dS) radius is $R = 10^{61} L_P$ .
If the c.c. is actually what it appears to be, no observation will ever probe distances larger than this.    The sphere of radius $R$ around us is called our cosmological horizon.

The only known symmetries, which can guarantee a vanishing c.c. in a gravitational effective field theory in four space-time dimensions are a combination of minimal Poincare supersymmetry\footnote{Of course, higher rank super-Poincare algebras can also force the c.c. to be zero, but the phenomenology of such models is not acceptable.} and a discrete R symmetry larger than $Z_2$\footnote{Continuous R symmetry is not allowed in $N=1$ SUGRA\cite{natizohar}.},  .  Generically, in  supergravity (SUGRA), SUSY alone implies a negative c.c.  .  However, in a conventional effective field theory approach, any attempt to correlate the small positive value of the c.c. with a small breaking of SUSY, seems doomed to failure.  Experiment tells us that the breaking of SUSY in the particle spectrum takes place at the scale of hundreds of GeV.  Dimensional analysis alone would then suggest a c.c. of order $\Lambda^{1/4} \sim 10^{10}$ GeV, which is $57$ orders of magnitude larger than the observational bound.  Calculations of Feynman diagrams in QU(antum) E(ffective) F(ield) T(heory), reproduce this value generically, though there are special models where somewhat smaller values can be obtained.   However, there is a subtlety in these calculations.  SUGRA contains an extra constant,$W_0$,  the value of the superpotential at the minimum of the effective potential, which is invisible in the $M_P\rightarrow\infty$ limit.   For any given size of SUSY breaking in the particle 
spectrum, we can tune $W_0$ to make the c.c. as small as we wish.  In QUEFT this appears to be ``unnatural fine tuning".  When the c.c. is (close to) zero, the gravitino mass is proportional to $W_0$.   This is because both $W_0$ and the gravitino mass, break any discrete complex R symmetry.

The present review is idiosyncratic.  It discusses some, but not all, attempts to understand how the small value of the c.c. is related to the breaking of SUSY.
Many such attempts are admitted by their authors to be failures.  It did not seem to be worthwhile discussing them.  I will concentrate on the following list of alternatives

\begin{itemize}

\item The String Landscape - Here one decides to separate the two problems from the outset.  The c.c. is assumed to be chosen, for anthropic reasons, at random, without regard to SUSY.  The question at issue is whether there is any reason to expect the SUSY breaking scale to be smaller than the Planck scale.  The answer depends on both the person answering, and the year.

\item Models in which our world is embedded as a non-supersymmetric defect in two large supersymmetric extra dimensions (SLED).  These are the most promising models for a non-anthropic solution of the c.c. problem based on SUSY, but rely on a size for the large dimensions, for which the low energy models present no dynamical explanation.

\item Ideas based on the Holographic Space-time formalism, and the related principle of Cosmological SUSY Breaking.  These ideas lead to a formula for the gravitino mass,
$$m_{3/2} = 10 K \Lambda^{1/4} ,$$ with $K$ nominally of order one.  This relation is barely compatible with current LHC data, and suggests very strongly that SUSY partners will be discovered in the next run of the LHC.

\end{itemize}

Let me expand a bit on the last item, since it is my own work.  The HST formalism can be motivated by a remarkable paper by Ted Jacobson\cite{ted}\footnote{... but historically it wasn't.  Fischler and I invented HST because of the work of Fischler, Susskind, and Bousso on the holographic principle\cite{fsb}.  It was only later that we realized the relevance of Jacobson's seminal work.} .  Jacobson's paper, which I will review below, showed that if one assumed an appropriate version of the Bekenstein-Hawking relation between area and entropy for an arbitrary causal diamond in space time, then Einstein's equations (but with the significant exception of the c.c. term), followed from the local version of the first law of thermodynamics: $dE = T dS$, in the limit of a trajectory whose Unruh temperature is infinite.  The entropy of an infinite temperature quantum system is the log of the dimension of the Hilbert space, so Jacobson's argument leads one to conjecture the Holographic Principle:  {\it The dimension of the Hilbert space describing all possible experiments in a given causal diamond is the exponential of one quarter of the area (in Planck units) of the maximal area surface on the boundary of the diamond, which is called the holographic screen.}   It also suggests that the gravitational field equations\footnote{In string theory, we have seen that all fields describe aspects of the geometry of compactified dimensions, so this statement applies to all the field equations of low energy physics.}  are the analogs of the hydrodynamic equations for a general quantum system.  The lesson from condensed matter physics, is that it only makes sense to quantize the hydrodynamic variables when we are describing low energy excitations of a system near its ground state.
In regimes of high entropy, {\it classical} hydrodynamic equations give a coarse grained description of the quantum dynamics of many systems, but have no relation to the underlying degrees of freedom.

The c.c. is an asymptotic boundary condition on this hydrodynamic description, which tells us how the asymptotics in proper time of a causal diamond is related to asymptotics in area/entropy.  For negative c.c., the area becomes infinite at 
finite proper time.  For positive c.c. area remains finite as proper time goes to infinity, while for vanishing c.c. the two go to infinity together, satisfying the relation $A \sim t^{d-2}$.  The maximally symmetric spaces for each value of the c.c. all have time-like asymptotic symmetry generators, and the c.c. also determines the asymptotics of the density of states at high energy.  

The HST formalism embodies Jacobson's ideas in a collection of real quantum models for various space-times.  These include early universe cosmology, through the end of inflation, a model of stable dS space, and the beginnings of a model for asymptotically flat space-time.  I'll outline salient points of the formalism below.

Both the String Landscape and SLED view the c.c. problem through the lens of effective field theory:  the c.c. is a calculable constant in a low energy effective Lagrangian, which is hyper-sensitive to short distance physics and has an experimental value wildly different from QUEFT estimates.  In the Landscape the cancellation is viewed as accidental, and primarily determined by environmental selection criteria.   In SLED it is a natural consequences of a certain type of large extra dimension scenario, but the explanation of the large dimensions is left for a more complete theoretical framework, and might end up having only an environmental explanation in that larger context.  

HST views the c.c. as an infrared boundary condition, which controls the number of quantum gravitational degrees of freedom that are not included in QUEFT.  It is not calculable but is chosen by hand.  A particular cosmological model, described below, allows for a distribution of long lived de Sitter (dS) universes, with varying c.c. and thus for environmental selection of the c.c. .  

This review is organized as follows.  In the next section I'll sketch the relation between SUSY, R symmetry and the c.c. in the context of effective $N=1$ SUGRA.  Then I'll review and criticize ideas about SUSY in the String Landscape.  The third section will review models of SUSY on a co-dimension 2 brane, which purport to achieve something close to the observed value of the c.c. ``naturally".  I'll conclude with an outline of the HST approach to the selection of the c.c., and its relation to SUSY breaking. 

\section{ $N = 1$ SUGRA}

$N = 1$ SUGRA has chiral multiplets, which contain complex scalar fields.  The potential for these fields has the form 
$$ V = e^K [K^{i\bar{j}} F_i \bar{F}_{\bar{j}} - 3 |W|^2] + f_{ab} D^a D^b.$$ In this formula, we've set the reduced Planck mass , $m_P = 2 \times 10^{18}$ GeV, to $1$, and  that will define the units for all of our equations.   $W$ is the superpotential, a gauge invariant holomorphic function of the scalars.The SUSY breaking order parameters are 
$$F_i = \partial_i W - K_i W, $$ and the $D^a$ terms of the various gauge groups.

Once non-perturbative constraints are taken into account\cite{natizohar} one finds that there are no solutions of the equations of motion with $F_i = 0$ but $D^a$ different from zero.  
At any rate, we learn from this equation that if SUSY is unbroken at the minimum of the potential, then that minimum is negative, with a value
$$ \Lambda = - 3 e^K |W|^2 ,$$ where the superpotential is evaluated at the minimum.   The maximally symmetric space with negative c.c. is $AdS_4$.  The isometry group of this space-time is $SO(2,3)$ .  We usually think of the SUSY generators of $N = 1$ SUGRA in 4 dimensions as a Majorana spinor .  The Dirac matrices are all imaginary in the Majorana representation, and therefore so is
$i \gamma_5$ which completes them to the Clifford algebra of $SO(2,3)$.  Thus there is a real representation of $SO(2,3)$, which acts on the spinor, made from commutators of these Clifford-Dirac matrices.  The product of two spinors is a sum of anti-symmetric tensors in general, and in this case the sixteen products are the sum of the anti-symmetric tensor of $SO(2,3)$ and a singlet.  We can make a consistent super-algebra if the anti-symmetric tensor is just composed of the generators of $SO(2,3)$ and the singlet vanishes.   

In the Majorana representation for the $SO(1,3)$ Dirac matrices, the spinors are real, so the anti-commutators of individual components are positive.  Consequently the Hilbert space of the theory must be a highest weight unitary representation of $SO(2,3)$.  In such representations, a generator conjugate to 
the rotation generator $H$ in the time-like two plane is positive.  This time-like direction appears to be periodic, but we can go to the universal cover of AdS space and make it an infinite line.  In the quantum theory, the real question is whether the spectrum of $H$ is quantized.   We would, at best, expect that if this model of quantum gravity on AdS space were integrable.   However, even in the infinite $N$ planar limit of maximally super-symmetric $SU(N)$ gauge theory, whose finite spectrum is completely integrable, the eigenvalues of $H$ depend on a continuous parameter and are not integers.  Thus, it appears that the proper geometry for models of quantum gravity is the universal cover of AdS, rather than AdS itself.

The penultimate sentence of the previous paragraph referred implicitly to the duality between models of quantum gravity on AdS space and conformally invariant field theories on the conformal boundary of $AdS_{d+1}$, which is a cylinder with topology $R \times S^{d-1}$, and a causal structure, but no metric.
The AdS/CFT correspondence will inform some of our considerations about the nature of the c.c. , but we do not have space here to discuss it in detail.  The canonical reference is \cite{aharonyetal}, and I will provide more references later, about specific points.  For the most part though, readers of this review will have to rely on my summary statements if they don't already know about the correspondence.

The algebraic discussion of the AdS SUSY algebra generalizes immediately to the case of De Sitter (dS) space, but contains a shocker.  The dS SUSY algebra exists, and its bosonic sub-algebra is $SO(1,d)$, which has NO highest weight unitary representations.  The spectra of boosts and rotations are both symmetric w.r.t. $0$.  Correspondingly, dS space has no everywhere time-like Killing vectors.  The Killing horizon of any generator that is time-like in a subregion bounds the causal diamond of any time-like complete trajectory.   That is the region that is causally accessible to a detector following that trajectory.  The subgroup of the dS group that preserves this diamond is $R \times SO(d-1)$ and it is possible to insist that the abelian generator is positive.  But there are no corresponding unbroken fermionic generators.  

These purely algebraic considerations confirm a fact that is obvious from the SUGRA formula for the potential:  no model of dS space-time can be supersymmetric.  SUSY is incompatible with positive c.c. .   A crucial phenomenological question then becomes, ``What is the functional relation between SUSY breaking and the c.c.  in the limit when either goes to zero?''.  The answers in the literature range from The Landscape  argument that there is {\it no} functional relationship (at least in some landscapers minds - M. Dine has tried to argue the contrary), to the field theory naturalness argument that
$\Lambda > M_{SUSY}^4$ where $M_{SUSY}$ is the scale of the largest splitting between superpartner masses, to my own contention that $m_{3/2} \sim \Lambda^{1/4}$ which means $M_{SUSY} \sim \Lambda^{1/8} M_P^{1/2} $ .

Another key question is ``Which came first, the $\Lambda$ or the $m_{3/2}?"$.  
Traditionally, one views the c.c. as a computed parameter in the low energy effective action of an underlying theory.  In AdS/CFT and HST, the c.c. is a primary quantity, which must be chosen in order to define the model.  SUSY breaking will be a consequence of the choice of positive c.c. \footnote{We'll discuss non-SUSY theories with negative c.c. a little, but will argue that models with vanishing c.c. MUST be SUSic.}.   Given this posture, we have to answer how nature chooses the c.c., and we'll see that, given a sufficiently rich cosmological model, the c.c. can take on a wide variety of values in different places in a large multiverse, with the ``choice" being simply a correlation between certain crude properties of our surroundings with the value of the c.c. .  The nature of this multiverse is very different in the String Landscape and HST models.

In order for the SUGRA Lagrangian to give vanishing c.c., we must either have $0 = F_i ,\ 0 = D^a \ $ and $ 0 = W$, or have a cancellation between the SUSY breaking terms in the potential and the term proportional to $|W|^2$.   The second possibility nominally seems to require a fine tuning.  To obtain a c.c. $\Lambda$ by choosing coefficients in the super-potential, Kahler potential, and gauge kinetic term, requires a tuning of order $$ \frac{\Lambda}{K^{i\bar{j}} F_i \bar{F}_{\bar{j}}} .$$  The experimental bounds on SUSY breaking give this as one part in $10^{59}$, for the observed value of the c.c. .   This is the essence of the claim that the c.c. is a fine tuning problem, akin to the tuning of the scale of 
weak interactions, in the standard model.  

I will review a class of models that have been proposed as a natural explanation of this tuning, but my own prejudice is that the c.c. is a very different object than the Higgs mass, because the theory of quantum gravity is not a quantum field theory.  Instead, the c.c. is a parameter that controls the density of states in the theory.  If it is negative, it controls the scale of the crossover regime from negative to positive specific heat for black holes, as well as the coefficient of $T^d$ in the free energy of the CFT dual to space-time.  When it is positive, it measures the dimension of the Hilbert space of the quantum theory.  This kind of parameter is not a coefficient in a low energy effective QFT action, but enters into the fundamental definition of the model.  It appears in the effective classical action because classical gravity is hydrodynamics, and the hydrodynamic equations need a large distance boundary condition.  

\section{The String Landscape}

The basic strategy of the String Landscape program is to exhibit a wide range of supersymmetric AdS compactifications, with randomly varying c.c. .   I will discuss it in a large class of models, related to those of KKLT\cite{kklt} .   Another large class of models, the large volume compactifications\cite{large} has similar properties, but I have not studied them thoroughly enough to review them properly.

One of the issues that must be addressed by this program is to obtain a compact manifold whose lowest Kaluza Klein energy scale is large compared to the inverse AdS radius.  A large class of models where there appear to be such classical solutions is defined by the rubric ``F-theory compactification on a Calabi-Yau 4-fold $(CY_4)$".   Ten dimensional Type II-B supergravity has the following bosonic fields

\begin{itemize}
\item A graviton $g_{\mu\nu}$
\item A complex scalar ``dilaton" $\tau$
\item A pair of two form fields $B_{\mu\nu}^i$
\item A four form field $C_{\mu\nu\lambda\kappa}$, whose field strength $F = dC$ is self dual $* F = F$.  
\end{itemize}

The covariant equations of motion for this model cannot be derived from a covariant action, but there are various gauge fixed actions from which they follow.  The equations are invariant under $SL(2,C)$ transformations on $\tau$, under which the $B_{\mu\nu}^i $ fields transform as a $[2]$, but this is broken to $SL(2,Z)$ by the Dirac quantization condition.  The models obtained for asymptotic values of $\tau$, in ten dimensional flat space, which are related by an $SL(2,Z)$ transformation are the same model.   This gauge equivalence is called S-duality.  

The basic solution that defines a class of F-theory models is one in which $g_{i\bar{j}} (z )$ is the metric on a 3 (complex) dimensional Kahler manifold and $\tau (z) $ is the complex structure of a family of tori of fixed area, fibered over that manifold, in such a way that the four complex dimensional fibre bundle is a Ricci-flat Kahler manifold\cite{vafaetal}, also called a Calabi-Yau manifold.   Calabi's conjecture, proved by Yau, shows that if the topological condition that the integral of the Kahler form of this four manifold, over every 2-cycle, vanishes,
then there is a solution to the Ricci flat equation.   In fact, there is a manifold of solutions, called a moduli space.   

$\tau(z)$ is not a function over the 3-manifold, but rather a section of a bundle, whose structure group is $SL(2,Z)$ .   That is, there are 4-cycles in the 3-manifold where $\tau$ is multivalued, undergoing an $SL(2,Z)$ transformation as $z$ circles the 4-cycle.  When the Minkowski dimensions of space-time are taken into account, these cycles are 7-branes in ten dimensional space.  These are not singularities of the $CY_4$ manifold, but only of its representation as a family of tori living on the complex 3-fold.  

The moduli of the solution are related to geometric properties of the manifold, volumes of p-cycles with $p=2,3$.   There is no modulus corresponding to the area of the torus.  More precisely, these solutions can be viewed as limits of 11 D supergravity compactified to $2+1$ dimensions on the $CY_4$ with variable area torus, in the limit that the torus area goes to zero.  In that limit, M2 branes, wrapped on the torus provide us with new continuous quantum number, which is interpreted as the momentum in the third spatial direction\footnote{This picturesque understanding of F-theory is fraught with difficulties.  Models of gravity in $2 + 1$ dimensional Minkowski space do not have S-matrix observables, since any scattering process creates a deficit angle in the geometry at infinity, which depends on the c.m. energy.  So it is not clear what object in the hypothetical $2 +1$ dimensional model converges to the F-theory S-matrix. The duality is probably valid for a non-compact $CY_4$.}.

We also have moduli corresponding to integrals of the $B$ fields over the $2$ cycles of the three fold.  All of the moduli combine together to form a complex moduli space with a Kahler structure\cite{BBS}.   In certain limits of the moduli space, one finds that $\tau (z)$ is constant, except at the location of the 7-branes. The base 3-manifold is then a singular $CY_3$ and the 
singularities are called orientifold 3 planes.  Nominally, the string coupling can be taken small everywhere outside the singularities.   However, in the presence of non-trivial $3$ form flux through the 3-cycles of the $CY_3$, there is a super-potential $W$ on the complex structure moduli space, which also depends on the parameter $\tau$ whose imaginary part is the inverse of the coupling.  
This fixes all of these parameters.  If the manifold has many three cycles, each of which can be threaded by various values of quantized flux, then we get of order $C^{B_3}$ different models, where $B_3$ is the third Betti number of the manifold.  The fact that $B_3$ can be as large as $500$ explains the wide variety of ``vacuum states" found by KKLT.  As we wander over this large collection of vacua we will find some for which ${\rm Im}\ \tau$ is large.

The value of $W$ at the stationary point $D_i W = 0$ is rarely zero, but there are points with discrete R-symmetry at which $W = 0$.  For these points, if $g_S$ is also small, we can construct a world sheet perturbation series, which describes a super-Poincare invariant $4$ dimensional universe.  KKLT instead look at points where $W_0$ is non-zero but small.  At zeroth order in perturbation theory, these are no-scale models, which are Poincare invariant but violate SUSY.
There are however divergences in the world sheet perturbation theory, due to a tadpole for $\rho$ the Kahler modulus of the $CY_3$.  The basic idea of KKLT is that these divergences are stabilized at point in Kahler moduli space, and give rise to a supersymmetric universe with negative c.c..  There is no known way to implement this idea in world sheet perturbation theory, so there is no real point in saying that the string coupling is weak.  We do not have any more reason to trust the KKLT construction for large ${\rm Im}\ \tau$ than we have for moderate values.
More details about the KKLT construction of flux vacua can be found in Chapter 10 of \cite{BBS}.  See also\cite{kd}.

The small parameter which justifies the approximations made in constructing the KKLT solutions, or more generally, F-theory solutions, is the inverse of the large volume of the 3-fold, in Planck units.  In the orientifold models, one also has the possibility of an everywhere small string coupling, but in fact there is no systematic world sheet construction of these models\cite{tb10to500}, so there is no useful sense in which this parameter controls the approximations.  Of course, a large volume is not a parameter in GR.  The volume must be shown to be large as some other parameter is tuned. That parameter turns out to be $W_0$, the value of the flux superpotential at the minimum of the potential.  We've already seen that if $W_0$ vanishes, we get a super-Poincare invariant flat space-time model.  For $W_0$ small, we can still solve the vanishing $F_{\rho}$ term condition.  Periodicity in the imaginary part of 
$\rho$, tells us that the leading large $\rho$ correction to the constant $W_0$ is of the form $ - A e^{-\rho / b}$.  If we use the leading large $\rho$ form of the Kahler potential the condition for SUSY becomes
$$a \frac{A}{b} e^{ - \rho / b} (\rho + \rho*) = 3 W_0,$$ which gives $|\rho |\approx - b {\rm ln}\ (|W_0|) ,$ self-consistently justifying the approximations if $W_0$ is exponentially small.
We get a supersymmetric AdS space, with a radius much larger than the volume of the compact manifold, which is nonetheless parametrically large.   The argument that $W_0$ can be small on the Landscape, echoes that of \cite{bp}:  $W_0$ is a complex number that gets contributions from all the different 3-cycles of the base, whose number is large.  There are likely to be points where it takes very small values.  It is only at these points that the analysis of KKLT is self consistent.   

Although it is far from rigorous, the analysis of KKLT to this point is robust and reasonable.  As we will discuss below, a truly rigorous argument for the existence of supersymmetric AdS models of quantum gravity would go through the AdS/CFT correspondence.  However, as a plausibility argument there is little to argue with.

The next step is where the KKLT argument goes into the realm of deep speculation.  KKLT argue that their models are most plausible for Calabi-Yau manifolds that are close to having a conical singularity and have {\it warped throats} where the metric is similar to a slice of Euclidean AdS space.  This leads to a possible exponential hierarchy of energy scales, in a familiar fashion.  The AdS/CFT correspondence suggests that this description is dual to a picture of D3-branes sitting at the singular point of the manifold.

Near a conical singularity, a manifold looks like
$$ds^2 = dr^2 + r^2 ds_{d-1}^2 .$$  Extending $r$ to run from zero to infinity, one obtains a non-compact manifold.   The theory of D-branes sitting near the singularity of such a non-compact Calabi-Yau manifold is well developed, and is, in supersymmetric cases, often dual to a quantum field theory.  In such a situation, one can break SUSY by adding anti-D-branes to the system.  The most sophisticated investigation of possible field theory duals to such a structure is\cite{dst}.  One does not expect it to be stable, but meta-stability is possible\cite{kvetal}.  

The basic idea of KKLT is that this local analysis is valid also for a compact Calabi-Yau manifold.  As far as I can tell there has been no plausible argument for making this extension.  The idea is, somehow, that if $W_0$ is small, and the manifold large, that one is making a small perturbation of the system.  One then calculates the shift in vacuum energy by adding the energy of the anti-branes, and obtains a positive c.c. .   This procedure is called {\it uplifting}, and all versions of the String Landscape depend on it.  

In my opinion, the claim that ``uplifting" is a small perturbation is manifestly false.  
The KKLT analysis of negative c.c. states, points, given our understanding of quantum AdS space, to a large family of super-conformal field theories.  Each point in the classical moduli space of F-theory solutions, to which the KKLT analysis applies, corresponds to a different SCFT.  The complete quantum data of each model consists of the correlation functions in that particular SCFT.  There are no quantum amplitudes for tunneling between models corresponding to different SCFTs.  Each has a unique stable vacuum, and localized fluctuations of AdS space, of sufficiently large size, all correspond to completely thermalized black holes.  Not only are there no expanding bubble instabilities, but there are no large meta-stable bubbles with another value of the moduli inside.  The different SCFTs all have very different asymptotic densities of states at {\it very high energy}.  

By contrast, the proposed theory for the ``uplifted" points is Eternal Inflation (EI).  EI is not a very well defined theory\footnote{We will explore the unique attempt to formulate a rigorous theory of EI below.}, but it certainly claims that all the different uplifted points in moduli space correspond to states in the same quantum theory, and that transitions between them constantly take place, providing a probability distribution that the world is sitting in each of them.  I fail to understand how this can be thought of as a small perturbation of $10^{500}$ {\it different} SCFTs, which do not communicate at all.

\subsection{Environmental Selection of Parameters}

The claim of String Landscape program is that there is a vast landscape of meta-stable dS states, which communicate primarily through CDL tunneling (see below). Local observations in one of these states are not supposed to interfere with local observations in another, so the prediction of the theory is supposed to be a probability distribution over possible universes (a multi-verse) with a variety of values for the c.c. .  The {\it a priori} probability distribution is generally assumed to be fairly smooth over the (discrete) landscape.   Indeed, given the exponential character of CDL tunneling rates, the only other option would be to say that one particular universe was picked with overwhelming probability, in which case the entire concept of a landscape would be irrelevant.  The smoothness assumption
sets up the conditions for anthropic/environmental selection of the c.c..

The anthropic principle was first posited on the basis of stellar nucleosynthesis calculations, which showed that the behavior of stars depended on rather fine tuned properties of nuclear levels.  Early discussions of it can be traced through the book of Barrow and Tipler\cite{BT}.  Although this sort of reasoning was applied to the c.c. very early, it is in principle incorrect to apply environmental selection criteria without the context of an actual model that generates different candidate universes in which the c.c. is a varying parameter.  Different theoretical models will give different probability distributions for the parameters that determine whether life can exist or not, and very few anthropic arguments are truly model independent. 

The first model of this type for the c.c. is due to Unwin and Davies\cite{UD}, and that was followed a few years later by independent work of Andrei Linde, A.D. Sakharov, and myself\cite{ALTB} .  When I presented this work in a UT Austin seminar, S. Weinberg remarked that since the c.c. was irrelevant to the evolution of stars and life once a galaxy formed, the anthropic bound was the dark matter density at the beginning of the matter dominated era, which is off by $12$ or so orders of magnitude\footnote{I'd suggested that it was of order the observational bound on the c.c., based mostly on ignorance of the details of galaxy formation. I've never quite forgiven myself for failing to investigate Weinberg's claim immediately.}.  It took two more years for him to realize his error:  galaxy formation arises from tiny fluctuations in the dark matter density, and it's the growth of these fluctuations that must compete against the accelerated expansion due to the c.c.\cite{weinbound} .

The success of the Weinberg bound, given the known values of the amplitude of primordial fluctuations and the dark matter density, is by now widely known, and the advocates of the String Landscape (landscapers from now on), cite this as a major success of the program.  In my opinion, anthropic arguments can only be evaluated within the context of specific models.  Any such model must provide us with a set of cosmologically meta-stable states and a description of which parameters of the low energy QUEFT vary independently over the landscape, as well as how the other parameters depend on those variables.   The {\it phenomenological} problem of the String Landscape is, that given our current knowledge, too many parameters appear to fluctuate.

Recall that the idea of the string landscape is to compute a potential on the moduli space of approximate supersymmetric solutions of the string equations of motion\footnote{They are approximate, just because of the existence of the potential.}.  We know that this moduli space contains points with many different gauge groups, matter representations and values of the physical Yukawa coupling which, in models that contain the $SU(1,2,3)$ gauge group with representations including the known elementary fermions, determine the masses and mixings of quarks a leptons, as well as a variety of couplings that could have been in the standard model, but are absent to very high accuracy\cite{bdg}.  These include couplings that lead to proton decay, neutron anti-neutron oscillation, lepton number and flavor violating interactions, quark flavor violating interactions, as well as the strong CP $\theta$ parameter.  

Extra discrete symmetries are rare in the String Landscape\cite{dz}, so one cannot justify the use of symmetry arguments to eliminate these terms, without proving that the probability distribution for meta-stable states is highly peaked on symmetric points.  No such argument is known.  Furthermore, as a solution to the c.c. problem, The String Landscape is overkill.  In the mid 1990s, when I worked on the String Landscape before it got its name\cite{tblandscape}, I advocated a scenario in which there was only one point in the landscape with c.c. small enough to satisfy some version of the anthropic bound.   If that were the case the rest of the low energy QUEFT would be determined by the anthropic argument fixing the c.c., and we could solve the problem of unexplained small couplings in the standard model by finding the unique point in moduli space with small enough c.c.\cite{tb95} .   However, the estimates of the number of points in the String Landscape is at least $10^{500}$, so there will be many points with c.c. small enough to fit the anthropic bounds.  There is no indication that insisting on even so severe a constraint as a low energy $SU(1,2,3)$ gauge theory with the standard model fermion content, reduces this number enough to find a unique point.

One is thus led to explore what other anthropic constraints on low energy physics exist.  The brief answer is: not nearly enough.  Given our lack of understanding of the mechanisms by which life arose, our complete ignorance of exo-biology, and the difficulties in working out the low energy physics of generic gauge theories, the problem of deciding which points in moduli space are likely to have living creatures in them, divides into two.  On the one hand, there are purely gravitational/thermodynamic arguments\cite{weinbergbousso}, which can constrain combinations of parameters like the c.c., the dark matter density, and the amplitude of primordial fluctuations.  These are, arguably, applicable to any conceivable form of life.   Any form of organized activity requires a relatively long period in the history of the universe in which localized entropy was small, so that the second law of thermodynamics could be operative.  The redistribution of energy in a system evolving from low to high entropy appears to be the only mechanism by which organized local activity could appear.    It's also reasonably plausible that galaxies are necessary for the existence of life.

Any arguments beyond that must simply assume that we are restricting attention to points in moduli space in which the low energy theory is more or less identical to the one we find experimentally, up to the energy scale that is probed by anthropic arguments.  We know so little about the chemistry of alternative gauge theories, and about the actual mechanism by which living things evolve from non-living matter, that we cannot explore strange regions of moduli space.  

Most authors assume that this restriction means that we must have the standard model gauge group and particle content, but that is not the case.   All anthropic arguments, of which I am aware, involve physics at scales no larger than the nuclear energy scale, up to perhaps hundreds of MeV.  The fine tunings of nuclear energy levels necessary to achieve stellar nucleosynthesis, which first gave rise to anthropic speculations, might convince us that we must have a theory which contained $SU(3) \times U(1)$ with up and down quark masses of roughly the order we observe.  The importance of radioactivity in the Earth's history, and in stellar processes, might convince us that we needed to have certain $4$ fermion interactions with the usual strength, but certainly not the full structure of the standard model.  Parity violating weak interactions are not required. Recall that it took $20$ years of experimental effort to make the approximate V - A form of weak four fermion couplings manifest.

It is therefore fair to say, that much of the standard model, and certainly the peculiar values of many of the parameters in the standard model, cannot have an anthropic explanation.  In string theory, extra generations of matter correspond to more tuning of moduli, small parameters do not appear to arise without a symmetry explanation, and symmetries are rare on moduli space.  

One cannot escape the implication that, on the basis of current theoretical knowledge, the String Landscape is ruled out by experiment.   It presents us with an array of possible low energy universes which is so vast that even the massive accident required to obtain a small enough c.c., leaves us with a class of models that is super-exponentially large.  Among these, the gauge group and representation content can vary wildly, but adding more massless species always involves tuning moduli to special points, so one would expect the typical model obeying only anthropic constraints to have only up and down quarks and a single electron and neutrino, an $SU(3)\times U(1)$ gauge theory with a confinement scale of order 100 - 200 MeV, and charge changing 4-fermi interactions with strength roughly equal to $G_F$.  These need not be of the canonical V-A form.  A single neutrino species could account for stellar cooling rates, and the behavior of supernovae might require a neutral current interaction of roughly the strength that occurs in nature, though one would have to explore the realm of general 4-fermi interactions more thoroughly than we have done before concluding this. 

It's important to remember that in assessing the Landscape predictions we are allowed to use only two criteria: is a meta-stable state centered at some point in moduli space of high probability in the {\it a priori} distribution, and does it give rise to low energy physics sufficiently similar to that we observe, that we can be fairly confident that life would exist.   There is no indication in the literature that the probability distribution strongly favors precisely three generations\footnote{Indeed, models with exactly three generations are rather rare.}, the chiral structure of the weak interactions, the many fine tuned parameters of the standard model (in the absence of symmetries), or the symmetries that might explain some of those tunings.  Instead, if the probability distribution is  
fairly uniform, then the underlying measure on moduli space strongly suppresses all of these choices, because they are concentrated on sub-manifolds 
of large co-dimension.

The situation is much worse if there are standard model super-partners at energy scales that could be relevant to the solution of the gauge hierarchy problem. The MSSM has many more operators of dimension $4$ and $5$, whose coefficients must be tuned severely in order to be compatible with experiment.  Thus, some Landscapers have tried to argue that the Landscape gives no correlation between the breaking of SUSY and the value of the c.c.\cite{douglassusskind} .  The solution of the gauge hierarchy problem is instead attributed to anthropic arguments\cite{anthropichiggs}.

On the other hand, Dine and collaborators\cite{dineetal} have given cogent arguments that the small c.c. and hierarchically small weak interaction scale {\it do} imply low SUSY breaking scale in the Landscape. Douglas\cite{douglasreview} also seems to have come to this conclusion.  

I'll try to outline these competing claims briefly.
The first argument is essentially counting\cite{douglassusskind}: Neglecting D terms the potential in SUGRA is given by (everything in Planck units)
$$V = \sum F_i \bar{F}^i - 3 |W|^2 . $$ The sum is over all of the moduli and the number of moduli is in the hundreds or larger. The superpotential has terms involving all of the moduli as well.  The anthropic constraint of very small positive c.c. says that the positive and negative terms in $V$ cancel to one part in $10^{123}$, but this is a single constrain on a very high dimensional space.  General considerations about functions on high dimensional spaces tell us that we should expect $V$ to have a number of stationary points exponential in the number of moduli.  Since the $F_i$ terms are in one to one holomorphic correspondence with the moduli and the moduli space is not compact, most of these stationary points will lie near the highest allowed values of $F_i$, which is to say that SUSY is typically broken at the Planck scale, the highest scale for which this effective field theory analysis could make sense.

The counterarguments have to do with meta-stability of the states.  SUSic points are absolutely stable, and one can argue that when the gravitino mass is much less than the Planck scale, their lifetime behaves like $e^{\frac{m_P^2}{m_{3/2}^2}}$.  By contrast SUSY violating points have no reason to be even locally stable.  The most detailed analysis of this is\cite{randomsugra}.  These authors show that the fraction of locally stable SUSY violating stationary points is of order $e^{ - c N^{1.5}}$, where $N$ is the number of moduli.  If SUSY breaking is small in Planck units, then this fraction increases to $e^{ - c^{\prime} N^{1.3}}$.  Even more interesting is the paper\cite{randomsugra2} which shows that the probability of obtaining a meta-stable dS minimum by ``uplifting" a supersymmetric AdS point is small unless $$N |W| < m_{SUSY}m_P^2,$$  where $m_{susy}$ is the scale of masses at the supersymmetric AdS point. This is because most supersymmetric AdS points have ``Breitenlohner Freedman allowed tachyons", fields with negative mass-squared, which are nonetheless stable in AdS space.
When this inequality is satisfied, the number of locally stable dS points, obtained by ``uplifting"
SUSic AdS stationary points is vastly larger than one would obtain from random dS stationary points.   Applying the above inequality to Kahler moduli in a KKLT model, we get $$m_S^{KK} \sim \frac{1}{\rho^p}   = ({\rm ln}\ W)^{-p} > N W .$$ This is a very mild inequality on $W$ for $N \sim 10^3$.  Indeed the plausibility of the KKLT analysis probably requires a much smaller value of $W$.  Complex structure moduli are even heavier.

The authors of \cite{dineetal} examined non-perturbative instabilities of locally stable dS stationary points. They argue that even though individual semi-classical tunneling probabilities are small, the vast number of possible final states in the Landscape would render many of these points unstable rather than meta-stable.  The individual tunneling amplitudes are suppressed if SUSY is broken at low energy\footnote{Actually there is a much wider class of non-supersymmetric stationary points which share this stability to tunneling.  If the scale of variation of $V$ as a function of the fields, is the Planck scale, then a finite fraction of all potentials is ``above the great divide"\cite{abj}, which means that their semi-classical lifetimes are of order the dS recurrence time.}, so this is another argument for low energy SUSY.
These authors conclude that the anthropic constraints on the weak scale, and these meta-stability considerations, imply that the majority of observable points on the Landscape have low energy SUSY.

\section{AdS/CFT}

As we've seen, the starting points for the KKLT construction of a String Landscape of meta-stable dS vacuum states are supersymmetric AdS vacua with c.c. that is parametrically small because of cancellations between of order $C^{B_3} $ positive and negative terms that are $o(1)$ in string or Planck units.  Here $C$ is of order one and $B_3$ is a Betti number, which can be of order $500$.  The analysis of Bousso and Polchinski\cite{bp} shows that there are rare vacua which have c.c. of order $C^{ - B_3}$, and the idea of the Landscape is that the dS vacua are all states of the same theory, and that anthropic selection will choose the rare, small c.c., states.   It is therefore somewhat disconcerting to realize that the AdS vacua from which the construction starts are definitely {\it not} states of the same theory.  The AdS/CFT correspondence gives us a rigorous definition of quantum gravity in $AdS_{d+1}$ space, in terms of a $d$ dimensional CFT.  In this correspondence, the value of the c.c. in Planck units is inversely proportional to a positive power of $c$, the ``number of fields" in the CFT.  $c$ is defined as the coefficient in the flat space free energy of the CFT.  
$$F = c K_0 V T^4,$$ where $K_0$ is the coefficient for a single massless scalar, $V$ the volume and $T$ the temperature.  This is a high energy property of the field theory, and the Wilsonian definition of field theory distinguishes models with different values of $c$ as having different quantum Hamiltonians.  $c$ is definitely not  a property one would imagine could be calculated in terms of a long wavelength bulk effective action.  

There has only been one proposal for finding the DOF responsible for large $c$.   Silverstein\cite{eva} pointed out that the counting of string junction states that could end on wrapped D-branes, scaled correctly with the Betti number, but there are several problems with this interpretation.   Usually, simple, weakly coupled physical states in string theory are associated with gauge invariant composite operators of low dimension in the boundary field theory, rather than with the underlying DOF.  The high temperature free energy is dominated by operators of very high dimension, which have dense spectrum of dimensions corresponding to the spectrum of a chaotic Hamiltonian\cite{LiuFestuccia}.  Furthermore, as we will see in a moment, a typical theory with a large number of almost free fields cannot have a c.c. which is small in {\it string} units.   There has been no followup on Silverstein's paper, attempting to verify her conjecture.

The problem of finding a c.c. small in string units is quite generic.  A typical CFT, in any number of dimensions, has a spectrum of primary operators of dimension $D$, which grows like the exponential of a power of $D$.  Each operator corresponds to a new bulk field whose mass is proportional to $D$ for large $D$.  By contrast, any model of a finite number of fields on $AdS_{d+1} \times K,$ with $K$ a compact manifold, has a mass spectrum which grows at most as a power of $D$.   Exponential growth is reminiscent of the stringy Hagedorn spectrum, so models in which the exponential growth sets in quickly can at best be interpreted as bulk theories with AdS radius of order the string scale. In order to have a regime in which SUGRA calculations are relevant to the CFT, we must have a large range of $D$ over which the exponential growth does not occur.  This is a necessary condition for having AdS radius much larger than the string scale.  
It is not a sufficient condition, despite the considerable amount of work that has gone into proving that it is\cite{sufficient}.  The counterexample is provided by large $n$ vector models with $d = 2,3$.  These appear to be dual to Vasiliev higher spin gauge theories in $AdS$, as long as we gauge the $O(n)$ or $U(n)$ or $Sp(n)$ global symmetry and insist that the gauge field strength vanish\cite{klebpolyaharony}\cite{rabO(N)}.  However, the higher spin indicates the existence of a leading Regge trajectory, and this is confirmed by the fact that when the theory is compactified on manifolds of topology more complicated than $S^{d-1}$, there are low energy excitations which can be identified as wrapped strings (and are indeed related to Wilson loops of the gauge theory).  The ratio between the string tension and the inverse AdS radius is not large for large $n$\cite{yinshenker}.  Large $n$ vector models with global symmetry groups {\it do have} an exponential density of higher dimension operators.
It would appear that any simple minded implementation of Silverstein's idea would be similar to these models.  

All known models with AdS radius much larger than the string scale are exactly supersymmetric (or are SUSY violating relevant perturbations of supersymmetric fixed points).   The large ratio is achieved by virtue of line or sequence of fixed points, with the parameter being related to the string coupling.  Large string coupling corresponds to large AdS radius.  Two methods have been proposed for generating SUSY violating deformations of these CFTs.  The first involves the concept of taking an orbifold of a CFT: projecting out states not invariant under a discrete symmetry group, and adding operators to the theory which are local w.r.t. to the invariant operators but non-local w.r.t. those that are projected out\cite{orbifold}.  This can be done in a way that preserves some or none or the original SUSY of the model.  Remarkably, whenever some SUSY is retained, the model still has large radius, but when SUSY is violated the radius jumps abruptly to the string scale.   A simple example is given by the orbifold of maximally SUSic SYM theory to a non-SUSic SYM theory.  To leading order in the 't Hooft expansion, the two theories have the same correlation functions for invariant operators.   However, at the next order in $1/N$, a beta function is generated for the 't Hooft coupling, and any fixed point would lie at $\lambda_{'t\ Hooft} \sim 1$, which is AdS radius of order the string scale\footnote{The situation is actually a bit more complicated because one must also include the RG flow of a marginally irrelevant double trace operator, which is forbidden by SUSY and affects the RG flow at tree level.  This complication does not change the conclusion that one cannot generate a SUSY violating fixed point with large radius AdS dual in this manner.}  

One can also try to find holographic RG flows\cite{holoRG} that lead to SUSY violating large radius fixed points, and/or violate SUSY only partially.  Quite remarkably one finds again that this exercise succeeds if some SUSY is retained but fails whenever SUSY is broken.  The candidates for SUSY violating fixed points always have tachyonic directions in the bulk scalar potential, which violate the Breitenlohner-Freedman bound\cite{BFbound}.  In terms of the AdS formula 
$$D - d = \frac{d}{2} - \frac{1}{2}\sqrt{d^2 + 4m^2 R^2} ,$$ relating the dimensions of operators to masses in AdS, these tachyons give rise to complex dimensions.

\section{Tunnels in the Sky}

The attempt to create a theory of the Landscape relies in an essential way on the Coleman-DeLuccia (CDL) theory of quantum gravitational tunneling.  This is a completely well defined problem in classical partial differential equations, whose application to theories of quantum gravity remains to be established.  I will briefly list a few results about the problem in PDEs, along with hypothetical implications of these for theories of QG.  We consider a potential for an arbitrary finite number of scalar fields, with a non-compact metric on field space, and a finite number of minima and maxima.  We will first discuss examples where the potential goes to infinity infinitely far away in the non-compact field space, and then the case relevant to the string Landscape, where the field space has finite volume and the potential vanishes at infinity.  We group potentials together into families which differ by an additive constant.

\begin{itemize}

\item For generic values of the additive constant none of the minima will be at zero c.c. .  If they are all positive, the CDL formalism defines a pair of transition probabilities for each pair of minima, which represent back and forth transitions between the two minima, and are related by the principle of detailed balance
$$P_{12} = P_{21} e^{S_2 - S_1} , $$ where the $S_i$ is the entropy of the $i$th dS minimum. The fact that the entropy rather than the free energy appears in this formula means that all of the states in the Hilbert space are being counted with equal weight and that the Hilbert space is finite dimensional.  This is of course just an interpretation of the semi-classical formula in terms of a hypothetical quantum theory of dS space, but it fits perfectly with the proposal we will make for that theory below.  The up-transitions, to the dS space with larger c.c., are interpreted as low entropy fluctuations of a system, most of whose states resemble the empty dS space with minimal c.c. .

\item Now consider the case with some positive and some negative minima, and tune the additive constant so that the lowest positive minimum has zero c.c.  .  We can ask whether the Lagrangian satisfies the positive energy theorem for the Minkowski solution.  For any potential $V$, we can consider the deformation
$$ V \rightarrow V + a \delta V,$$ where $a$ is a positive constant and $\delta V$ a non-negative smooth function, which vanishes outside a small neighborhood of the regions where $V$ is negative.  It's clear that for large enough $a$ the positive energy theorem is valid, and examples show that the same remains true in some cases where the full potential has negative minima.  Thus, the space of potentials with a Minkowski solution and some negative minima, satisfying the positive energy theorem has co-dimension one in the space of all potentials.  If we now add a small positive constant to the potential, the resulting dS minimum does have a tunneling solution which dumps the system into the basin of attraction of some of the negative c.c. minima.  However, the tunneling probability behaves as $e^{-S}$, where $S$ is the dS entropy, and the tunneling process is consistent with being a low entropy fluctuation.  The semi-classical calculation is consistent with a model of a stable quantum system, with a finite number of states, most of which resemble empty dS space.  The fluctuation does not produce an AdS vacuum state, but rather a singular Big Crunch.  The causal diamonds of segments of time-like trajectories inside the classical geometry, which evolves from the instanton, have finite area holographic screens, and the areas are generically much smaller than that of the largest dS horizon.

\item When there is no positive energy theorem at the Minkowski minimum, no one has proposed a consistent quantum interpretation of the instanton.  It certainly appears to be an instability, but there is no proposal for the stable state to which it decays.  Furthermore, excitations of the unstable vacuum, modeled by black holes, are not excitations in the decaying region.  Indeed the instanton does not penetrate into the black hole interiors.   The final state of a decay from a generic excited state is a geometry with multiple singular regions, only one of which resembles the interior of the expanding bubble of instanton.

\item A final issue for the Landscape interpretation of CDL instantons from dS minima, is the fact that Euclidean dS space is compact.  As a consequence, the dilute gas approximation breaks down unless the number of instantons is $\ll (R_{dS} / R_I )^d $, where $R_I$ is the size of the region over which the instanton differs from the false dS solution.  For potentials ``above the great divide"\cite{abj}, the interpretation of this bound is simple.  The physics of such systems is consistent with a Hilbert space with a finite number of states, which can be causally accessed by a single geodesic.  The finite number of instantons reflects the finite number of localized tunneling events accessible to this trajectory before the entire system thermalizes.   It is not consistent with the Eternal Inflation picture of an infinite number of tunneling transitions in a universe of ever expanding spatial volume.

\item If all of the minima are negative, we are in the realm of the AdS/CFT correspondence.  Near the false vacuum $AdS_{d+1}$ space, the linearized equations of motion have two solutions for each direction in scalar field space.   The instanton is a one dimensional path in the space of fields, so finite action solutions asymptote to some particular direction in field space.  The AdS/CFT dictionary identifies configurations with only the more rapidly falling power behavior as representing a state in the Hilbert space, while the one which falls less rapidly, or grows, represents a perturbation of the theory by some operator.   The powers are 
$$\lambda_{pm} = \frac{d}{2} \pm \frac{1}{2}\sqrt{d^2 + 4m^2 R^2} .$$  Here $R$ is the false vacuum AdS radius and $\frac{m^2}{2}$ is the coefficient of the quadratic term in the action for fluctuations around the minimum.  $d$ is the dimension of the boundary of AdS, on which the CFT lives.     Positive $m^2$ is dual, via the AdS/CFT correspondence, to irrelevant perturbations.  The analog of Coleman's ``shooting" argument for non-gravitational instantons shows that we can tune the coefficient of the growing solution to zero by the choice of the value of the field at the center of the instanton (the point at which $\frac{d\phi}{dr} = 0$), and the instanton has collective coordinates.   The details have been discussed in \cite{harlowrabbarbon}.  The bottom line of the analysis is that, because of the growth of the volume of AdS at infinity, instantons pile up at the boundary, which is dual to the UV of a putative dual field theory.  There is no sense in which a local observer in AdS experiences a meta-stable vacuum.   It is likely that there are no models of QG corresponding to effective field theories possessing such instantons.

\item When $m^2 < 0$, there is a one parameter set of instantons corresponding to the perturbation of the CFT by a relevant operator, with the parameter being the relevant coupling in AdS radius units\footnote{In Poincare coordinates, the instanton is a domain wall, and corresponds to RG flow.  In global coordinates, the RG flow is cut off by the finite spatial volume.}.  The instantons have {\it no} collective coordinates, since their asymptotic behavior always contains the non-normalizable mode.  The instantons are homogeneous on $dS_d$ sections of $AdS_{d+1}$.  The dual boundary field theory interpretation is a relevant perturbation of a CFT on $dS_d$.  In \cite{barbrabharlsuss}, the singularity of the instanton on the AdS boundary is interpreted as the coordinate singularity of the $dS_d$ slicing.  An interesting point made by Maldacena\cite{maldaadstunnel} is that all of these instantons have two different interpretations, since they can be viewed as satisfying either the Dirichlet boundary conditions on the non-normalizable solution\cite{maldaadstunnel} or a boundary condition relating the normalizable to non-normalizable mode.   The first of these is a perturbation of the CFT by a relevant ``single trace" operator.   The second perturbs the CFT by a monomial function of the single trace operator\cite{HH}.  The distinction between the two models can only be seen once one decides how to quantize the fluctuations around the instanton solution, but at the quantum level the distinction is enormous. For example, in the interpretation of \cite{HH}, the instanton corresponds to a quantum Hamiltonian which is unbounded from below, and might only be defined by   imposing a cutoff and adding a ``dangerous irrelevant operator" to the Hamiltonian.
In the interpretation of \cite{maldaadstunnel}, the perturbation is relevant and bounded from below when the field theory is quantized on $dS_d$.  In neither case does the instanton signal the decay of a meta-stable vacuum to a stable one.

\item In Section 2, I complained about the absence of a non-perturbative definition of the hypothetical meta-stable uplifted dS states of the String Landscape.  The FRW/CFT correspondence\cite{frwcft} is an attempt to provide such a definition.  
The formalism is motivated by the semi-classical picture of the String Landscape.  One assumes a field space isometric, at large values of the fields to the moduli space of some 4 dimensional compactification of (ten or eleven dimensional)  SUGRA, with a potential $V$ that vanishes along the non-compact directions of moduli space, where SUGRA formally becomes exactly supersymmetric in ten or eleven dimensions, with vanishing c.c..  In addition to these asymptotic ``minima", the potential has minima at positive and negative values of $V$.
Now consider an instanton mediating between some $dS_4$ minimum, and the basin of attraction of the asymptotic region.  The geometry of the analytically continued instanton is an FRW, and it has an $SO(1,2)$ isometry group.  The hypothesis of FRW/CFT is that there is a conformally invariant theory on the boundary 2-sphere whose gauge invariant correlators define the theory of the String Landscape.   The hypothetical theory is very different from AdS/CFT because the boundary metric is a fluctuating variable, described by a Liouville theory with a Liouville coordinate with negative (time-like) signature.  Gauge invariant correlators are analogous to integrated vertex operators in string theory.   One of the first questions one asks about such a theory is how it depends on the choice of instanton.  There are two aspects to this:  which meta-stable dS minimum do we choose, and to which SUSic asymptotic region does it decay?  The answer that the proponents of this theory want to believe is that {\it it doesn't matter}, and the argument for this depends on the idea of bubble collisions, and chains of CDL instantons.  It is certainly possible, though by no means guaranteed, that on a complicated potential landscape, dS to dS transitions would eventually lead from any dS minimum to any other.  Let's consider the last dS minimum along some chain before decay into the SUSic zero c.c. . The usual field theoretic treatment of multiple instantons in dS space (though {\it not}, as we've seen, the analysis of actual Euclidean solutions) leads one to expect an infinite number of collisions  between this last bubble and other CDL bubbles nucleated very close to the light cone to which the bubble asymptotes.  In FRW/CFT these are interpreted as contributions to the correlation functions of the boundary theory.  On the other hand, one can view the same collision from the point of view of any bubble chain that ends in the same penultimate dS space, and decays into any SUSic basin of attraction.  One can see the vague outlines (though, in my limited view only the vague outlines) of an argument that all SUSic basins of attraction, and all penultimate dS spaces lead to the same boundary CFT.  A serious problem with this point of view is the observation\cite{hfs} that collisions between two bubbles whose terminal points are in the basins of attraction of two different SUSic asymptotic regions create horizons in the FRW.  No single time-like trajectory has causal access to the full asymptotic two sphere, on which the boundary theory actually lives.   This is particularly vexing because the expectation is that there are an infinite number of such collisions for any bubble.

\item  Another crucial issue for FRW/CFT, on which there has been almost no work, is the quantum meaning of the integrated CFT correlators.  If we interpret the same two dimensional Euclidean Lagrangian as a perturbative string background, then the genus zero correlators do not define a self consistent quantum theory.  The theory must be second quantized, which leads one to an infinite sequence of higher genus corrections to the correlators.  This is a very general property of any theory defined by a 
path integral over fields, one of which (in this case, presumably, the Liouville field) provides the time coordinate, so that surfaces of fixed value of this field define states in the Hilbert space.   One cannot avoid configurations in the path integral on which the fixed value surfaces are disconnected.  Unitarity then forces one to introduce higher genus surfaces in the path integral.  In conventional string theory, the growth of the volume of higher genus moduli space makes the sum over genera ill defined.  This problem was addressed in \cite{shenkeretal} and arguments, whose validity I am not able to assess, were advanced to show that the corresponding series in FRW/CFT is convergent.  As far as I know, no proposal has been made to define the Hilbert space of the quantum theory defined by FRW/CFT, the mathematical meaning of the correlation functions as amplitudes, or their connection to any observations made in our own universe.

\item A final point that has never been addressed by the proponents of FRW/CFT is the treatment of the expanding bubbles of zero c.c. space-time as effectively four dimensional, even though the moduli flow asymptotically to a decompactified ten or eleven dimensional region.  This is crucial to the proposal.  The boundary geometry of a ten or eleven dimensional FRW has eight or nine dimensions and a Euclidean field theory of metric fluctuations does not exist above two or perhaps three dimensions.   It is certainly true that at any finite time, one can view the semi-classical geometry is a theory of an infinite number of fields in four dimensions, but one would expect a dramatic effect of the asymptotic pileup of an infinite number of massless four dimensional modes.  No discussion of this problem appears in the FRW/CFT literature.

\end{itemize}

My personal conclusion from all of this analysis, is that the theory of CDL tunneling provides no positive support for, and lots of negative evidence against, the proposal of a String Landscape.  There are two situations in which there is a clear quantum mechanical interpretation of CDL tunneling.  The first and more hypothetical of these is the interpretation of CDL transitions for potentials with dS minima ``above the Great Divide"\cite{abj} in terms of a finite dimensional Hilbert space with entropy equal to one quarter of the area of the largest radius dS minimum.  Empty dS space is a thermodynamic description (see the discussion of Jacobson's principle in the section on Holographic Space-time) of the generic ensemble of states. 
The transitions from that minimum are temporary and improbable excursions into meta-stable low entropy states.  In the case of transitions to negative c.c. crunches, one cannot calculate the reverse transition probability using the CDL formula, but the principle of detailed balance tells us what it is.   The other case that is fairly well understood is that of instantons dual to perturbations of a CFT on a boundary dS space.  In the case where the perturbation is (at least marginally) relevant, and the dS Hamiltonian is bounded from below, the boundary field theory provides a rigorous definition of the instanton transition.  It is not really a transition or decay, and the instanton has no collective coordinates.  The nature of the boundary field theory depends crucially on the boundary conditions imposed on the fluctuations around the instanton.  In at least one case, one of the two natural choices leads to a theory which makes sense only in the presence of a cut-off, and an irrelevant operator that bounds the Hamiltonian from below at large boundary field strength.  Neither of these two situations fits with the expectations of Eternal Inflation or a Landscape.   Instanton solutions which have collective coordinates in AdS space, lead to the conclusion that the AdS solution is not meta-stable in any sense, and decays on a time scale of order its radius.

For unstable Minkowski solutions (more generally, dS minima below the Great Divide), the CDL analysis leads to enormous confusion.  Black hole excitations of the ``meta-stable vacuum" do not become excitations of the singular final state of the ``decay".  The instanton does not penetrate into the black hole interior, and in the semi-classical limit, for black holes whose size is not exponential in the semi-classical expansion parameter, the singularity of the interior geometry of the black hole is reached long before a bubble can nucleate inside the black hole.  There can also be universe saving conspiracies, in which the decay is prevented by collapse of a black hole around the expanding bubble.  To make a long story short, we have absolutely no understanding of what the quantum theory might be that corresponds to these instantons.

The only case of CDL tunneling, which might fit the Landscape paradigm, is the situation with asymptotically SUSic regions of field space, which is the basis for the FRW/CFT proposal.  Partisans of that approach may be encouraged by the fact that many other cases of CDL tunneling seem to be meaningless.  I am more impressed by the myriad conceptual challenges that I've outlined.   When combined with the phenomenological challenges I presented in Section 2, I conclude that the String Landscape is an hypothesis of dubious validity.

\section{Supersymmetric Large Extra Dimensions}

Supersymmetry is aesthetically appealing and seems to be required for consistent string models of quantum gravity, but much of its cachet has come from the fact that it seems to solve the hierarchy problem of the standard model, and could explain why the weak scale is so much lower than the Planck or unification scales.  Much of the recent disillusion with SUSY is a consequence of the fact that most extant SUSY models suffer from a mild degree of fine tuning, once experimental constraints from the LHC are taken into account.  

Nonetheless, even in the late 1990s, while SUSY was still an object of admiration, phenomenologists took up another idea from string theory, and 
proposed that the apparent disparity between the weak and Planck scales was an illusion caused by the fact that the world we perceive is a 3-brane\footnote{The world volume of a 3 brane is a 4 dimensional space-time.} embedded in a $4 + D$ dimensional space-time\footnote{There are actually two versions of this, one in which $D = 1$ and the world is a thin sandwich of $AdS_5$ bounded between two branes\cite{Randallsundrum}, and the other in which the warping of the extra dimensions is not severe\cite{LED}. The work we are reviewing does not incorporate the Randall-Sundrum idea.}  .  In classical gravity, compactification on a space of volume $V$ leads to a parametric relation
$m_P^2 = V m_D^{D + 2}$.  The idea was to take $m_D \sim 1-10$ TeV and convert the hierarchy problem into the question of why the compact volume is so much larger than the Planck scale.  These models are thus called Large Extra Dimension or LED models.

Another idea that was tried out by various people was that LED might explain why the brane we live on is so close to having a flat space-time geometry, despite the fact that known bounds on supersymmetric particle energies suggest that radiative corrections to the c.c. are large.   Let me illustrate this with a simple model\footnote{In this section I will be following the crystal clear Les Houches lecture notes of C. Burgess\cite{burgessoncc}, who is one of the originators of the idea of SLED.}.  Consider the Einstein-Maxwell Action in $6$ space-time dimensions, coupled to two 3-branes: 
$$S = - \int\ d^6 x \sqrt{-g}\ (m_6^4 R + \frac{1}{4} F_{MN}F^{MN} + \Lambda_6 - \sum_{b = 1,2} \int\ d^4 x \sqrt{- \gamma} T_b .$$ $\Lambda_6$ is the 6D c.c. and the $T_b$ are the brane tensions. $\gamma$ is the metric induced on the branes by $g$.   We look for solutions that are warped products of a 4D symmetric space and a two dimensional compact internal space, within which the 3-branes are point singularities.  
$$ds^2 = e^{2 W(y)} g_{\mu\nu} (x) dx^{\mu} dx^{\nu} + g^{(2)}_{mn} dy^m dy^n .$$
We postulate a non-zero flux of $F_{MN}$ in the internal directions $$F_{mn} = f \epsilon_{nm} , $$ a scalar function with a non-zero, quantized integral over the internal manifold
$$ \int\ \sqrt{g_2} f = \frac{2\pi n}{e_6}.$$ $e_6$ is the coupling of the Maxwell field.  This stabilizes the extra dimension and prevents it from collapsing.

If the brane tensions are equal, the two dimensional metric is a sphere with two conical deficits at the brane positions
$$g_{mn} dy^m dy^n = L^2 (d\theta^2 + \beta^2 \sin^2 (\theta ) d\phi^2) .$$
$$F_{\theta\phi} = Q \beta L^2 \sin (\theta ),$$
and $1 - \beta = \frac{m_6^4 T}{2\pi}$.  The constants, including the constant scalar curvature, $R$, of the 4 dimensional space-time are:
$$Q = \frac{n}{2\beta e_6 L^2} , \ \ \ R = m_6^4 (Q^2 - 2 \Lambda_6) \ \ \ L^{-2} = [ 1 \pm \sqrt{1 - \frac{3n^2 m_6^4 \Lambda_6}{8 \beta^2 e_6^2}} ] . $$
We see that without tuning $\Lambda_6$ we do not get a vanishing or naturally small curvature on the branes.

In $6$ or more dimensions, supergravity (SUGRA) forbids a six dimensional cosmological constant in the presence of branes of co-dimension $2$ or more.
Furthermore, we will see eventually that the scale of the extra dimensions will have to be $ \sim 100 (eV)^{-1} $, and phenomenological constraints\cite{addbdn} forbid more than two extra dimensions from being this large.
The creators of this scenario called it Supersymmetric Large Extra Dimensions (SLED)\cite{burg262728}.

The physics on the brane need not be supersymmetric, and the SLED enthusiasts chose it to be just the standard model.  The basic idea of these models that if the extra dimensions are large enough, the presence of a localized non-supersymmetric brane should not modify the existence of a stable flat brane configuration in 6D Minkowski space.
 Because the extra dimensions will be quite large (of order the final vacuum  energy, $1/L\sim  10^{-2}$  eV), we must choose six dimensions in order to be consistent with the fact that the extra dimensions have not yet been detected.
 A non-supersymmetric brane can be consistently coupled to 6D SUGRA by using a brane localized non-linear realization of SUSY.  There need be no standard model super-partners at all, and the partisans of SLED consider the failure to find the MSSM at the LHC to be a phenomenological success of their model.  Of course, if super-partners are discovered tomorrow, we could modify the model to account for them, so it is perhaps better to say that the extant models are part of a family of models with 4D super-partners at a variety of energy scales, and that LHC data constrains the super-partner masses.
 
 There  are  a variety  of supergravities  from which  to choose in six dimensions,  but one choice is particularly convenient in that it allows a rugby ball solution, very similar to the one described  above.  This supergravity is the Nishino-Sezgin chiral gauged supergravity , whose rugby-ball  solutions  are  described  by the following 6D bosonic fields:  the metric,  $g^{MN}$ , a scalar  dilaton,  $\phi$,  and  a specific $U (1)_R$   gauge  potential,  $A_M$ .   To  lowest  orders in  the derivative   expansions,  the (supersymmetric)  bulk  and  (non-supersymmetric)  brane
contributions to the action are

$$S  = - \int d^6 x\ \sqrt{-g}\frac{1}{2\kappa_6^2}g^{MN}  (R_{MN} + \partial_M \phi \partial_N \phi) + \frac{1}{4} e^{-\phi} F_{MN} F^{MN}  + \frac{2e_6^2}{\kappa_6^4} e^{\phi} - \sum_b \int\ d^4x\ \sqrt{ - \gamma} (T_b - A_b \epsilon^{mn} F_{mn} ).$$
 $M,N$ are $6D$ indices, $m,n$ 2D indices, and
$A_b$   can  be interpreted  as describing  the amount of Maxwell flux, $\Phi_b = A_b e^\phi /2\pi$, that is localized on the brane.
The simplest situation is when the two branes are identical, in which case there is a rugby ball solution to these field equations of the form described above, but  with $ W  = 0 $ and the 2D metric and field strength replaced  by
$$g_{mn} dx^m dx^n = l^2 (d\theta^2 + \beta^2 \sin^2 \theta d\chi^2) e^{-\phi_0 } ,$$	$$F_{\theta\chi}  = Q\beta l^2 e{-\phi_0} \sin \theta .$$ $\phi_0$ is the constant value of $\phi$ in the solution. It is not determined by the field equations.
The radius of the extra dimensions is $L = l e^{- \phi_0 /2}$, with $$ L^2 e^{\phi_0} = (\frac{\kappa_6^2}{4 e_6^2}) .$$
  The  Ôflat directionÕ $\phi_0$ is a consequence of a classical scale invariance  of extra-dimensional super- gravity,  under  which $g_{MN}  \rightarrow \zeta^2 g_{MN}$    and  $\phi \rightarrow \phi - 2{\rm ln}\zeta$ .  This symmetry is preserved  by the branes  only if $T_b$ is $\phi$-independent and
$A_b  \propto e^{-\phi}$ , and  is broken  otherwise.   When  broken,  the brane  interactions  lift  the flat
 direction and give a mass to the corresponding  KK mode.  In the case of most interest for the cosmological constant both  $T_b$  and  $A_b$ are $\phi$-independent,  and  so the breaking of the scaling symmetry by the $A_b$  term fixes $\phi_0$  via the flux-quantization condition
$$\int_{M_2} F + \frac{1}{2\pi} \sum_b A_b e^{\phi_0} = \pm \frac{1}{e_6} .$$
The brane  metric is flat for any choice of brane  Lagrangian,
and in particular regardless  of the value of $T_b$ .  It can be shown that this follows quite generally when $T_b$ and $A_b$ are $\phi$ independent (note this is not the choice which preserves the classical scale invariance of the bulk action)\cite{burgessoncc}.  It is crucial to take brane back reaction into account, to obtain this result.  It is also necessary that the stress energy off the brane vanish.
 
 The  existence  of extra-dimensional  solutions  that allow flat  4D geometries  to coexist  with large 4D-lorentz-invariant energy densities does not  in itself solve the cosmological constant problem.  For different choices of brane properties there are also other solutions for which the on-brane  geometries are curved.  One must re-ask the cosmological constant question in the
6D context:  first identify which features  of the branes  are required  for flat brane  geometries, and then ask whether these choices are stable against integrating out high-energy  degrees of freedom.
The  two necessary  conditions for flat branes are precisely the kinds of criteria that can be preserved  by loops of the heavy Standard Model particles,  because these reside on the branes.
For example,  having large tensions makes the branes  like to be straight and so not bend into the off-brane directions.  So any particles trapped on such branes  will have stress energy only in the on-brane directions:  $T_{mN}  = 0$. On-brane  loops do not make the branes more likely to bend, because they typically do not reduce the brane  tension.
As far as the second criterion is concerned, since there  are  initially  no couplings  between  any  on-brane  particles  and  the bulk  dilaton,  such  couplings  are  never  generated  by doing only loops of on-brane  particles but require bulk loops as well.
 These  arguments are borne  out by explicit calculations  \cite{burgess45}. Quantum corrections  that also involve the extra-dimensional bulk  fields are potentially dangerous,  because they can generate  brane-$\phi$ couplings even if these are initially absent. The brane  does couple to the metric  and bulk  gauge field and  these  couple to the dilaton,  so loops involving  the metric  and  gauge field can generate nonzero couplings between branes  and $\phi$, and thus a c.c. on the brane.
Explicit  calculations  \cite{burgess4630} show that supersymmetric cancellations allow the effective  4D
 vacuum  energy to be of order  the Kaluza-Klein  scale, $L^{-4}$.
 The non-canceling  bulk quantum contributions are naturally very small, because an N -loop contribution is always proportional to a factor of $e^{2N \phi_0}$, which is a very small number.  This is essentially the suppression of bulk loops by powers of the inverse volume of the extra dimensions.   The one loop calculation gives
 $$\Lambda = \frac{k}{16 \pi^2 L^4} ,$$ and higher loops are suppressed by higher powers of $L$ in Planck units. $k$ is a number of order $1$.

 In certain  circumstances  the result  can  even be smaller  than this  because  the branes  sometimes  fail to break  supersymmetry  at  one loop \cite{burgess30}.   In  this  case the leading quantum  contribution  to the observed  4D cosmological  constant  turns out to involve  one
brane  and one bulk loop, and is smaller by a factor of $\frac{1}{16 \pi^2}$.
 
These estimates can fit the observed c.c., if $L$ is large enough. There  are two phenomenological upper  limits for L. These are tests of NewtonÕs  inverse-square  law and astrophysical constraints coming from cooling of supernovae and red giant stars by emission of bulk Kaluza-Klein modes.   In the conventions used by Burgess, these presently give an upper  limit $ L < 45 \mu$m, and so $1/L > 4 \times 10^{-3}$  eV.  For reasonable values of the $o(1)$ constants, this is consistent with the observed vacuum energy because of the powers of $16 \pi^2$.
Astrophysical  energy-loss  constrains  the extra-dimensional  gravity  coupling,  $\kappa_6  <   (10 {\rm TeV})^{-2}$ .  The relation between the 4D and 6D gravitational couplings is $\kappa_6^2 \sim 4\pi L^2 \kappa^s$, so $L < 1.3 \mu$m . Assuming we are right at the bound gives the intriguing limits $\Lambda (1 loop) \sim (0.04 {\rm eV})^4 $ and $\Lambda (2 loop)  \sim (0.01 {\rm eV})^4$ , which, given the cavalier treatment of order-unity factors, is very close to the observed density.  A more careful treatment of the defect angles at the branes seems to give even better results\cite{burgessoncc}.
 
 The SLED scenario predicts large extra dimensions close to but perhaps not within reach of current experiments and could motivate a variety of new attempts to search for these, both at the LHC, and in tests of gravity at short distances.  It is advertised as predicting the absence of standard model super-partners at the LHC, but I believe this is a choice of parameters in a wider class of SLED models.  If super-partners are discovered in the next few years, I suspect that we will see that wider class of models.  I also suspect that concrete implementations of the SLED models, with calculable parameters may have a version of the little hierarchy problem.  Quadratic divergences in the effective potential of the Higgs field are cut off around the 6D Planck scale, but that scale is considerably higher than the Higgs or Z mass.  Nonetheless, the SLED scenario has the virtue of making fairly concrete predictions that we will see lots of interesting new physics at energy and distance scales that are perhaps within reach in the near term.  
 
 The biggest problem, in my view, with SLED is that it provides no dynamical explanation of why the extra dimensions are large.   Absent such a mechanism, one is forced to consider anthropic explanations, and that would require a complete understanding of the UV completion of the model.  Are the SLED models just points in the String Landscape (in which case many of the criticisms of the previous sections apply)?   If not, what is the proper context for understanding the choices made in these models?   SLED is an attempt at bottom up, low energy effective theory, model building.  Its explanation of both the value of the c.c. and the gauge hierarchy problem relies on an assumption of large extra dimensions, but as far as I can see it does not provide a low energy calculation showing that the dimensions have the size required by phenomenology.   It does seem to provide an economical way to solve multiple problems of high energy physics, but its score in the ``competition" to explain the puzzle of the c.c. is hard to assess without greater understanding of the underlying theory of quantum gravity.
 
  \section{Holographic Space-time, SUSY, and the Cosmological Constant}

The basic ideas of Holographic Space-time (HST) are best explained by starting from a remarkable paper by Jacobson\cite{ted}.  Jacobson considers an arbitrary point $X$ in a Lorentzian space-time and a small causal diamond whose holographic screen intersects that point.  A causal diamond is the intersection of the interior of the backward light-cone of a point P and that of the forward light cone of a point Q in the causal past of P (so that there is a time-like trajectory going from Q to P).
The boundary of the causal diamond is a pair of light cones, each of which have
a metric $$ds^2 = g_{ij}^{\pm} (x,u) dx^i dx^j .$$ $u$ is a null coordinate on the boundary, and the holographic screen is the value of $u$ where the $d-2$ volume of the Euclidean metrics $g_{ij}^{\pm}$ is maximum.  It may lie on the backward or forward light cone, or as in Minkowski space, on their intersection.

For Jacobson's argument, we take a causal diamond for which the point $X$ lies on the intersection of the light cones, and whose size is so small that the curvature of the space-time is almost negligible.  We assume that the space-time curvature radius is much larger than the Planck length, so that the area of the holographic screen of the diamond can be much larger than the Planck area.

Jacobson considers an accelerated trajectory, which, near X, has an Unruh temperature that goes to infinity.  He defines the expectation value of the Hamiltonian for this trajectory as
$$E = \langle H \rangle = \int\ T_{\mu\nu} (x(\tau ) k^{\mu} k^{\nu} d \tau,$$ where 
$k^{\mu}$ is the tangent vector to the trajectory, which becomes null in the limit of infinite temperature.  $\tau $ is the proper time.  The integral includes the integral over an area ${\cal A}$ on the holographic screen transverse to the trajectory. The expectation value is taken over all quantum states accessible in this causal diamond, since we are taking the infinite temperature limit.  The energy goes to infinity in the limit, because of the time dilation as the tangent vector becomes null. This infinity will cancel against the infinite temperature in the first law of thermodynamics
$$ dE = T dS .$$   

Now invoke the Bekenstein Hawking formula for entropy
$$ S = \frac{\cal A}{4 L_P^{d-2}}, $$ and use Raychauduri's formula for the change in area

$$ d\theta /d\lambda  = - R_{ab}k^ak^b, $$, where $\theta$ is the expansion\footnote{The expansion is defined by $$ \delta A = \int_{\cal H} \theta d\lambda dA ,$$ , where $\delta A$ is the change in area transverse to a pencil of generators of the Rindler horizon.} or a congruence of null geodesics nearly parallel to the nearly null Unruh trajectory, $k^a$ is the tangent vector to the trajectory, and we have neglected terms that are small for small causal diamonds in the vicinity of the tip of the local light cone.

There are two important facts about this formula.  The first was pointed out by Jacobson.  We have derived Einstein's equations as the hydrodynamic equations of any (quantum) system that obeys the quantum mechanical Bekenstein Hawking relation between entropy and area.  In typical situations, the fields appearing in the equations of hydrodynamics are classical fields defined in terms of averages over a large number of quantum variables which can be in a doubly exponentially large set of states.  In other words, the entropy of the system is macroscopic, and that is of course the case if we use the Bekenstein Hawking formula for an area large in Planck units.

String theorists will immediately object that we {\it know} that the gravitational field is quantized, but they forget that all the models they are used to are models of small excitations of systems near their ground state.  It is well known that one does quantize small fluctuations of hydrodynamic fields around the ground state.  These are known as phonons in condensed matter physics.  They exist even when the excitations of the system far from the ground state are completely different.  It would be a mistake to model the high temperature quantum behavior of copper, at {\it e.g.} its melting point, in terms of the low energy sound waves that characterize its low temperature behavior.   The classical equations of hydrodynamics remain valid in all temperature regimes, though the constants appearing in these equations can change.

The other remarkable fact about Jacobson's argument is that the cosmological constant is not determined by local hydrodynamics.  This is not a bug but a feature, and it shows the error of associating the c.c. with local zero point fluctuations of quantized fields.   We will see that in fact the c.c. is a large scale boundary condition, determining the nature of the asymptotic Hilbert space of causal diamonds with infinite proper time separation between their tips.

\subsection{The Covariant Entropy Principle and Holographic Space-time}

't Hooft\cite{'t Hooft} was probably the first person to suggest associating degrees of freedom with a black hole horizon, and we will see that his idea about the BH entropy counting the shapes of the horizon was remarkably prescient.  However, it was important to divorce these ideas from the specific context of black holes, and also to understand the way in which they were related to particle physics.  This has been a long and arduous path, and is still not complete, but it began with a seminal paper of Fischler and Susskind\cite{fs}, which was generalized by Bousso\cite{bousso} to a principle valid for arbitrary causal diamonds in arbitrary space-times.  

Bousso's version of the principle can be stated simply for causal diamonds, though it applies to more general regions of space-time.  It is simply that the entropy of the maximally uncertain density matrix in a quantum theory describing a causal diamond, is one quarter of the area, in Planck units, of the holographic screen of the diamond.  This entropy is the natural logarithm of the dimension of the Hilbert space.  We will often refer to it as the entropy of the Hilbert space or the entropy of the diamond.   

This principle led immediately to the conjecture, by Fischler and myself\cite{tbwf}, that the value of the positive cosmological constant in an asymptotically de Sitter universe, was simply a measure of the finite dimension of the space of quantum states in such a universe.  Indeed, the maximum area of causal diamonds in such a space-time is $4\pi (RM_P)^2 ,$ where $R$ is the de Sitter radius.  $R$ also determines the mass of the maximal mass black hole that is allowed in such a space-time, whose Schwarzschild radius is $\frac{2R}{3}$.  I've written these formulae only for four dimensions, because I believe that is the only dimension in which a quantum theory of dS space makes sense.

In space-times of negative c.c., asymptotic to anti-de Sitter ($AdS_d$) space, the c.c. determines the point in proper time where the causal diamond of a geodesic trajectory has an infinite area holographic screen.  The causal diamonds of larger proper time contain a time-like segment of the conformal boundary ($R \times S^{d-2}$) of $AdS_d$, and the quantum mechanics of propagation in that boundary time is a quantum field theory.  One indication that this is the case is that the spectrum of black hole states of very high energy in $AdS_d$ is
given by the entropy formula
$$ S(E) = c_d (ER)^{\frac{d-2}{d-1}} ,$$
where $$ c_d = \frac{A_{d-2}}{4} (\frac{R K_d}{L_P})^{\frac{d - 2}{d - 1}} ,$$ $A_{d-2}$ is the area of the unit $d-2$ sphere, and $K_d$ is the constant multiplying $L_P^{d-2}$ in the $d$ dimensional Newtonian potential.  This is the formula for the asymptotic density of states of a $d - 1$ dimensional quantum field theory on a $d-2$ sphere of radius $R$.  Notice that $R$ controls what would be the number of free fields in the UV fixed point CFT, if it were free.  This parameter is always a measure of the number of high energy DOF, even in an interacting CFT.

In both these cases the c.c. controls the high energy behavior of the theory.  It cannot be a parameter determined by the vacuum of a quantum field theory, which is a low energy choice if there are multiple vacua.   This is perhaps the strongest of many indications\cite{tbgadfly} that ideas about a Landscape of vacua, which guide the considerations of the previous two sections, are incorrect.

In the case of $\Lambda = 0$, for space-times which are asymptotically flat,  the area of a causal diamond goes to infinity with the proper time, according to the formula $ A \sim t^{d-2}$.  

\subsection{Holographic Space-time}

These kinematic considerations suggest a complete quantum mechanical formulation of geometry.  A time-like trajectory through space-time defines a nested sequence of causal diamonds, corresponding to some nested set of proper time intervals.  For a cosmological space-time we would choose intervals $[0, t_n]$, where zero is on the Big Bang hypersurface, while for a time symmetric space-time it's more convenient to choose $[- t_n , t_n ]$ .  It turns out that if we want to keep manifest rotation invariance at all times, the intervals have to be chosen to be the Planck time, that is $t_n - t_{n - 1} = L_P$.  We'll measure everything in Planck units from now on.  

If $t_n > t_m$, then the $n$th causal diamond contains points space-like w.r.t. the entire causal diamond at $t_m$, so we should view the Hilbert space of the smaller causal diamond as a tensor factor in that of the larger one.  In other words, a time-like trajectory is equivalent to a nested tensor factored Hilbert space
$$ {\cal H}_n = {\cal H}_{n-1} \otimes {\cal N}_n .$$ We'll see how to choose the relative sizes of the two factors when we discuss the nature of the variables.

The dynamics of the system must respect this tensor factorization.  That is, the Hamiltonian {\it must} be time dependent.  It operates in a Hilbert space ${\cal H}_{\infty}$ infinity, whose entropy is determined by the maximal size causal diamond
encountered as $t_n \rightarrow \infty$, but splits into $H(t) = H_{in} (t_n) + H_{out} (t_n)$, where $H_{in} (t_n)$ operates in ${\cal H}_n$ and $H_{out} (t_n) $ in its tensor complement in ${\cal H}_{\infty}$\footnote{To be more precise, the operator algebra in the full Hilbert space factorizes as $A_{in} \otimes A_{in}^{\prime}$, where the prime refers to the commutant of $A_{in}$ in the full operator algebra. $H_{in\ (out)}$ is built from operators in $A_{in\ (out)}$.}.  If the latter is infinite dimensional, we have to be a bit careful about taking the limit.   Readers fond of time translation invariance should recall that in general relativity, global symmetries are properties of the asymptotic boundary of space-time and should not be manifest in a local formulation.

In principle, the description of physics along one time-like trajectory suffices for a complete description of the universe, but we like to talk to our friends, so a good theory should include separate quantum systems for every time-like trajectory in space-time, with mutual constraints guaranteeing a consistent description of shared information.  A sufficiently rich set of trajectories would then determine everything else.  Note that such a formulation solves the notorious {\it problem of time} in quantum gravity.  Time, in HST, is multi-fingered.  The difference between HST and the Wheeler DeWitt formulation is that we do not attempt to incorporate all of the information about all possible time-like trajectories, into a single Hilbert space.  This accords with the reality of causality restricted measurements.  At any given time\footnote{A synchronization of the proper times of different trajectories is implicit in the phrase {\it at any given time}.  That synchronization depends on the space-time geometry and a natural one has been found in all cases studied so far.}, the overlap between the causal diamonds of two trajectories
misses regions of space-time causally accessible to one and not the other.

The quantum version of this is that for each time, and each pair of trajectories ${\bf x , y}$, we introduce an overlap Hilbert space ${\cal O} (t_n , {\bf x, y})$ which is a tensor factor in both ${\cal H}_n ({\bf x})$ and ${\cal H}_n ({\bf y})$.  The dynamics and initial conditions along each trajectory prescribe two different density matrices in this overlap tensor factor and we must have
$$\rho ({\bf x} , t_n) = U^{\dagger} ({\bf x , y}, t_n) \rho ({\bf y} , t_n) U ({\bf x , y} t_n ) .$$ There are further constraints for multiple overlaps, but in the simple models we have solved they are all satisfied.

To make sure that we have a sufficient number of trajectories, we give the space of trajectory labels the topology of a Cauchy surface in space-time.   In all the models we have studied the topology is that of $R^{d-1}$ and we've made the space of labels into a cubic lattice.  The geometry of the lattice is unimportant and it's clear that any other regular lattice will do.   It's likely that we could be more abstract and label the trajectories by the zero simplices of any simplicial complex with the simplicial homology of $R^{d-1}$ but we have not explored this.

We always choose the synchronization of proper times so that the Hilbert spaces ${\cal H}_n ({\bf x})$ have the same dimension.  We choose the overlap for nearest neighbor points on the lattice, to have the same entropy deficit relative to the individual ${\cal H}_n$ as the entropy deficit of ${\cal H}_{n-1}$ in ${\cal H}_n$. We choose the latter, in rotation invariant situations, to be such that the time difference is one Planck time, in which case nearest neighbor trajectories are always one Planck distance apart at any time.  The rest of the overlaps are chosen in one of two ways, which we believe to be equivalent:

\begin{itemize}

\item We can search directly for systems of overlap conditions which satisfy the HST consistency condition.  This was the way we worked in constructing an HST model of the very earliest moments of a Big Bang universe.  It turned out that these models described FRW universes\cite{holocosm}.

\item Using Jacobson's dictum that Einstein's equations are the hydrodynamic equations for a quantum gravitational system, obeying the Bekenstein-Hawking area law, we can choose a solution of Einstein's equations and take the geometrical overlap conditions of that space-time as a guide to the overlap conditions for the quantum theory.  We followed this path in constructing theories of Minkowski space and effective theories of black holes.

\end{itemize}

The claim of equivalence is partly obvious, if one takes Jacobson's point of view about space-time.  We have a quantum theory in which the correct area/entropy law, and the correct causality relations are built in to the definition of space-time in terms of quantum mechanics.  The space-time that emerges from this definition, in the limit that we look only at causal diamonds of large area, is, at least in all the cases we've studied, smooth, and Jacobson's considerations apply.  Thus, it satisfies Einstein's equations with a stress tensor derived by looking at the Hamiltonians of local Rindler trajectories\cite{holounruh}, matching our quantum prescription for this trajectory to Jacobson's relation between the stress tensor and the expectation value of the energy in the collection of maximally accelerated local Rindler frames.

This does not however mean that {\it every} valid solution of Einstein's equations 
corresponds to an HST model.  There are solutions that violate the covariant entropy bound.  There are asymptotically flat solutions without supersymmetry.
There are solutions in dimensions so high that supersymmetry implies that there can be no long range gravitational interaction.  On a more subtle level, there are de Sitter solutions in dimensions other than $4$.  We will see below that local physics in de Sitter space is approximately super-Poincare invariant, and that every quantum theory of de Sitter space has a discrete parameter, which can be tuned to the super-Poincare invariant limit.  
Four dimensions is the unique case in which one can write a classical supergravity that has de Sitter solutions, which can be tuned to obtain a model with zero c.c. and a scattering matrix.

\subsection{The Quantum Theories of Minkowski and de Sitter Space}

In the interests of Space and Time, I will present the theories of Minkowski and stable eternal de Sitter space only in four space-time dimensions. I will also present them in a hypothetical model in which the only low energy particles are the minimal SUGRA multiplet.  

The theory of scattering in a gravitational theory in Minkowski space is best formulated by viewing the scattering states as a representation of the super-BMS algebra\cite{awadagibbonsshawtb}.  In four dimensions, this algebra has the form

$$ [\psi_{alpha}^{\pm} (P) , \bar{\psi}_{\dot{\beta}}^{\pm} (Q) ]_+ = \sigma^{\mu}_{\alpha\dot{\beta}} {\rm min}\ (P , Q )_{\mu} \delta (P \cdot Q) .$$
The awkward notation ${\rm min}\ (P , Q )_{\mu} $ must be explained as follows. $P$ and $Q$ are null momenta, positive (outgoing) energy at future null infinity and negative (outgoing) energy at past null infinity.  The delta function says that they are parallel, and then one of them is a smaller positive multiple of $(\pm 1, {\bf \Omega} ) $, with ${\bf \Omega}$ a unit 3-vector.  A more elegant notation assigns a projector $E_P$ in the unique hyper-finite Type $II_{\infty}$ factor\cite{connes11D}, to each null momentum and writes the minimum as ${\rm Tr}\ E_P E_Q$.

In Minkowski space-time, conformal infinity is not a manifold.  It consists of two cones, and the scattering matrix maps a Hilbert space defined on the past cone 
to one defined on the future cone.  At each non-singular point of either cone, there is a unique outgoing null-vector (positive energy for the future cone and negative energy for the past) $P$, and we parametrize the operator algebra by these null momenta.  At the tip of each cone, there is a degeneracy in the choice of $P$, even when one has fixed the angle ${\bf Omega}$, and the mass of stable massive particles parametrizes the proper linear combination of the two outgoing momenta that point along ${\bf \Omega}$.  We will not discuss massive particles in this review.

The anti-commutation relations are accompanied by the compatible constraint,
$$ \sigma^{\mu} P_{\mu} \psi = 0,$$ and its Hermitian conjugate.  We can then write the $\psi (P)$ in terms of chiral conformal Killing spinors $q_{\alpha} (\Omega )$ on the two-sphere
$$\psi_{\alpha} (P) = \int\ d\Omega_0 \ \psi (\Omega , \Omega_0) q(\Omega_0 ) ,$$ where
$$ \psi (\Omega , \Omega_0) = \sum Y_A^{*i} (\Omega_0 ) \psi_i^A ,$$ where $Y_i^A$ are the complete set of chiral spinor spherical harmonics and
$$[\psi_i^A , \psi_B^{\dagger\ j}]_+ = \delta_i^j \delta_B^A .$$ We've parametrized the spherical harmonics by writing the sum of all half integer spins as the product of the $N$ dimensional representation of $SU(2)$, with the $N + 1$ dimensional representation, in the limit $N \rightarrow \infty$.  

The $\psi_{\alpha} (P)$ are operator valued measures on the future null cone in momentum space.  That is, for every measurable $f^{\alpha } (P)$ they define an operator $\psi [f]$.  In order to describe scattering states we must impose a constraint $$ \psi [f] | Jet \rangle = 0, $$ unless the support of $f$ for non-zero momenta is concentrated in a finite number of disjoint spherical caps.  I use the terminology ``Jets" to indicate the relation to the infrared finite QCD observables defined by Sterman and Weinberg.   I believe the formulation of gravitational scattering theory in terms of the super-BMS algebra will lead to IR finite amplitudes, perhaps related to those defined by Fadeev and Kulish\cite{fadkul}.
Particles are clearly singular configurations of the super-BMS generators.
It is worth noting that particle statistics (the fact that permuting particles is a gauge symmetry) follows automatically from our realization of particles as a localized excitation of the super-BMS fields on the sphere and the correct connection between spin and statistics (as well as the $( - 1)F$ gauge symmetry responsible for Fermi statistics) follows from the algebra.

The definition of jet states does {\it not} imply that the $\psi_{\alpha} (P)$ annihilate the state at most angles.  $\psi [f]$ can be non-zero outside the opening angles of the jets, if $f$ is concentrated at $P = 0$.   When we formulate this more precisely by taking $N$ finite, we will see that $\psi [f]$ must truly vanish in small annuli around each spherical cap, but need only be concentrated at zero momenta over the bulk of the sphere.  We refer to all the variables outside of the jets as {\it horizon DOF}.  Those in the annuli around the jets are {\it frozen horizon DOF}, while those which are non-vanishing, but carry zero momentum are called {\it active horizon DOF}.  The formulation of scattering in terms of the super-BMS algebra thus suggests the possibility that the real theory of Minkowski space has more in it than particles.

HST allows us to formulate this somewhat singular discussion as a limit of something more precise.  The holographic principle implies that a finite causal diamond, whose holographic screen has area $4\pi R^2$, has a Hilbert space whose entropy is $\pi (RM_P)^2$.  We can implement this cutoff on the variables $\psi_i^A$ by simply keeping $N$ finite.  $N$ is then proportional to $R M_P$.  
The finite diamond analog of the jet constraint is

$$ \psi_i^A | Jet \rangle = 0,$$ for of order $NK + Q$ matrix elements, where $ 1 \ll K,Q \ll N$.  We are used to thinking of the indices $i,A$ as running over the values of the $z$ component of angular momentum, but the commutation relations are invariant under $U(N) \times U(N + 1)$ transformations on $\psi_i^A$, so we can also think of more localized bases on the sphere.  

The constraints imply that the matrix $M_i^j  \equiv \psi_i^A \psi^{\dagger\ j}_A$, is block diagonal with ``small" blocks satisfying 
$$\sum K_i = K ; \ \ \ \sum K_i K_j = Q ,$$ and one large block of size $N - K$.
We identify the small blocks with elements of a basis of the space of spinors on the sphere, which are localized near a point ${\bf \Omega_i}$.  The matrix elements that vanish on the jet state are identified with super-BMS generators concentrated in annuli around the jets, and the large block represents the active horizon variables.

In each such sector of the Hilbert space, we write a time dependent (that is, $n$ dependent) Hamiltonian
in a causal diamond 
$$ H_{in} (n) = H_{in} ( - n) = \sum P_0^i + \frac{1}{n^2} \sum_{r=1}^L g_r (n) {\rm tr}\ M^r ,$$ where the coefficients $g_r (n)$ approach constants at large $n$ and $L$ is a fixed integer $ > 7$.  It does not grow with $n$\footnote{In higher dimensions, the order of the polynomial grows like $n^{d-4}$.}. Note that, according to 't Hooft scaling, the second term is a small perturbation of the first when $n$ is large.

We take the $P_0^i$ to be the momenta appearing in the Super-BMS algebra localized at angle ${\bf \Omega_i}$.   These are constructed in a uniform way and the scale of the momentum is proportional to $K_i$.  Thus, for large $n$ all of these Hamiltonians propagate particles freely.  It is easy to see that they all conserve the quantum number $K$, which is the sum of the particle energies, asymptotically.  This is a consequence of several things.  Firstly, in the Hamiltonian $H_{in} (n)$ with $n \ll N$, operators which could remove $o(N)$ of the constraints simply do not appear.  Secondly, because of the finite order of the polynomial, a single action of the Hamiltonian at $n \sim N$ cannot remove $o(N)$ of the constraints.  Finally, the $1/n$ relative to 't Hooft scaling prevents the interaction part Hamiltonian from acting repetitively at large times.   This means that $Q$ will change, so that the number and momenta of individual jets of particles will change from the incoming to outgoing states, but their total energy is conserved.
So, as might have been expected in a theory meant to approximate general relativity, our proposal for a quantum theory of Minkowski space, there is no notion of energy locally, which is either conserved or trajectory independent, but there is an asymptotic conservation law.  We will see in a moment that the axioms of HST show that the Hamiltonian must be built in such a way that the $K$ is also trajectory independent.  

Before doing that, we expand a little more on the nature of particle interactions in these models, and the role of the Hamiltonian $H_{out} (n)$.  As the causal diamond of our trajectory (call it Trajectory $T_1$) gets smaller, coming in from the infinite past, its $H_{in}$ may no longer contain the DOF corresponding to some of the particles in the incoming asymptotic state.  By following which of the blocks are contained in this trajectory's ``in" tensor factor through a sequence of times $n_r$, we can track the particles in space-time.  Note that, as long as all the $n_r$ are large, the constraints have a robustness, with small quantum fluctuations, so these tracks are real markers of the quantum state of the system.  

Blocks which are no longer included in the causal diamond of $T_1$, but will be contained in the causal diamond of some other trajectory $T_2$ during this sequence of times.  The overlap conditions between the two trajectories at early times\footnote{And between a whole sequence of intermediate trajectories, which interpolate between the two even at times when they each have small causal diamonds.}  force $H_{out} $ of $T_1$ to have a copy of the dynamics 
and the states that are described by $H_{in}$ of $T_2$.  

Thus, the full quantum state of ${\it any}$ of the trajectories, including both {\it in} and {\it out} tensor factors, contains information about all of the particle tracks in the incoming state.   Particle identification begins to break down when the inequality $n_r \gg \sum K_i$ (where the sum runs only over those particle jets that are included in the causal diamond), fails.  {\it Parametrically, this is just the criterion that the size of the causal diamond be larger than the Schwarzschild radius of the c.m. energy.}  Although we do not have time to demonstrate it here, this parallel works in any number of dimensions\cite{holonewton}.  This means that, as we go back to larger causal diamonds in the future, the number and energies of particle tracks leaving the causal diamond with large enough $n_r$ that the number of constraints is large, will have changed.  From the point of view of a coarse grained theory with time and length scales $\gg n_r$, this will look like a particle interaction.  

Another possibility, if the $n_r$ at which particles lose their identity is large enough, is that {\it no} particle degrees of freedom will emerge immediately from
the small causal diamond.  This is black hole formation.  The correct energy/entropy relation for black holes is built in to the HST formalism, so if the interaction Hamiltonian is a fast scrambler\cite{sekinosusskind} the system will have the properties expected for black holes from semi-classical considerations.
In particular, if the system is following a generic path through the state space of a Hilbert space with entropy $n_r^2$, then the probability of being in a state satisfying of order $K_i n_r$ constraints is $e^{- K_i n_r}$ and so the hole will emit particles at the thermal rate predicted by Hawking.  

Because the $H_{out}$ of each trajectory contains copies of the particle interactions seen in the $H_{in}$ of other trajectories, the outgoing particles from some interaction can affect the initial state of the system in some other scattering event.   Thus, we can begin to see how the Feynman diagram description of particle interactions will emerge from the HST formalism.  By following the patterns of constraints on the $\psi_i^A$, which change via the action of the interaction Hamiltonian, and imposing the HST constraints, we can expose a global space-time picture, which has a crude resemblance to a Feynman diagram.  Causal diamonds where particle interactions occur are connected together by freely propagating particles.   Yet the formalism also has the flexibility to describe the formation and evaporation of black holes, quantum mechanically.   In addition, as shown in \cite{holonewton} it gives rise to the correct long range eikonal scattering of relativistic particles at large impact parameter, which, by abuse of language, we call Newton's Law.  The significance of this is that, since our Hilbert space contains only transverse gravitons\footnote{The representation theory of a delta function localized super-BMS generator is precisely that of a light-front superparticle.}, the long range tail of scattering does not come from Feynman diagrams.  Instead, it turns out that the active horizon DOF, and the assumption that their state is generic, are the guarantors of the correct scaling of the large impact parameter amplitude.

For most choices of the polynomial in the interaction Hamiltonian, the scattering matrix defined by HST will not be Lorentz invariant or conserve spatial momentum.  However, the consistency conditions of HST {\it require it to satisfy these symmetries}.  Consider two time-like trajectories in Minkowski space, at relative rest and separated by a distance $d$.  The rules of HST require us to describe the quantum mechanics along each of these trajectories, in such away that all of the density matrices for overlapping information are unitarily equivalent.  

As the proper time $N$ in each trajectory's causal diamond goes to infinity, the entropy deficit of the overlap causal diamond, relative to either of the individual diamonds, is $Nd$, which is a declining fraction of the total entropy in the limit.  If the state in each diamond's Hilbert space were generic, this would immediately imply, by Page's theorem\cite{page}, that the overlap state is completely entangled with its tensor complement in both of the individual diamonds, which means that the full states in the two Hilbert spaces must be identical.  

The lack of genericity implied by the definition of particles does not disturb this argument, because, in the quite concrete sense described above, particles leave tracks.   A particle in the causal diamond of $T_1$, but not in that of $T_2$, will, at large enough $N$, left a track in the holographic screens of a sequence of $T_2$'s diamonds, by imposing constraints.   Thus, the information about particle DOF is certainly contained in the overlap and the unitary equivalence of the asymptotic evolution of any state is guaranteed.  Similar remarks apply to two trajectories related by a Lorentz boost.  Thus, the coefficients $g_r (n)$ should be chosen in such a way that the unitary representation of the super-Poincare group in the Hilbert space of the super-BMS algebra commutes with the S matrix defined by the Hamiltonian.

To summarize this brief review of the HST theory of Minkowski space: we have proposed a collection of Hamiltonians, and a definition of particle (more properly, jet) states, which describe particle scattering, in a manner reminiscent of Feynman diagrams, but also describe the possibility of black hole formation.   The theory contains a representation of the super-BMS algebra in the limit of infinite causal diamonds, and we have argued that the unitary representation of the full super-Poincare group on this representation space commutes with the S-matrix, if we really implement all of the overlap constraints of HST.  

\subsubsection{The Quantum Theory of Stable dS Space}

To leading order in inverse powers of the radius of curvature of dS space, our model for a time-like geodesic in dS space, is the same as our model for Minkowski space, except that after a proper time of order the dS radius, when the Hilbert space acted on by $H_{in} (n)$ has entropy equal to the dS entropy, we continue to evolve in proper time with the time independent Hamiltonian 
$$H_{dS}  = \sum P_0^i + \frac{1}{R_{dS}^2 } \sum g_r {\rm tr}\ M^r .$$ Proper time goes on forever but the Hilbert space ceases to expand.  This is of course the message of dS geometry, and was the basis for the conjectures of \cite{tbwf}.

The Jet states are now defined by a constraint on the initial conditions at time $ - R_{dS}$.   If we imposed them at time $- \infty$ we would quickly evolve to equilibrium.  Thus, the theory of stable dS space is artificial, in the sense that it's initial conditions, at a time of order $R_{dS}$ in the past are special and require explanation.  The actual quantum model, with generic initial conditions in the remote past, evolves to equilibrium fairly quickly.  

Nonetheless, with the understanding that we incur a debt to explain the initial conditions of cosmology,
the model with the above $H_{in} (t) = H$ for $|t| > R_{dS}$ and particle constraints on the initial conditions at $t = - R_{dS}$, followed by Minkowski like evolution in the interval $ [- R_{dS} , R_{dS}]$ , is a plausible model for particle physics in our university, which is consistent with all classical and semi-classical data on the quantities that could be measured along a time-like trajectory in our own universe.   

That is, there is a period where particles scatter with only minor modifications to their amplitudes in Minkowski space, followed by a period of thermalization, in which more and more of the local degrees of freedom are absorbed by the thermal bath.  The natural time scale for thermalization with the Hamiltonian
above is $R_{dS}$, but many local excitations can have much longer lifetimes.  
The $R_{dS}$ scale is associated with geodesic motion of localized excitations unbound to the trajectory, out through the cosmological horizon.  Stable and meta-stable excitations localized on the trajectory have much longer lifetimes.
For example, the lightest charged particle can decay only via the nucleation of its anti-particle from the thermal bath, after which they annihilate and the decay products propagate out to the horizon.  The time scale for this process is $m^{-1} e^{2\pi m R_{dS}} + R_{dS},$ (where $m$ is the particle mass) and is typically dominated by the first term.  Similarly, black holes following the trajectory whose Hamiltonian we are describing will survive for a time $R_S^3 + R_{dS}$ before they thermalize completely.  We do not yet have an HST description of gauge forces\footnote{This depends on a proper understanding of compactification of extra dimensions in HST, a subject that is only in its infancy.}, so we cannot yet verify the first of these claims, but the second follows from the description of black hole physics above.

The ``minor modifications of particle amplitudes" described above includes supersymmetry breaking.  The mechanism for SUSY breaking is the absorption and emission of gravitinos from the horizon\cite{susyhor}.  Since there are no chiral mirrors, the effect of this is to induce a gravitino mass. I described this originally in terms of Feynman diagrams :  Consider a self consistent Feynman diagram calculation of the gravitino mass, analogous to a gap equation for chiral symmetry breaking.  We evaluate a contribution to the gravitino effective action coming from diagrams where two virtual gravitino lines propagate to the horizon, where a vertex operator $V$ emits and absorbs them.   The amplitude is
(to leading exponential order in $R_{dS}$) 
$$ e^{- 2 m_{3/2} R_{dS}} \sum V^{\dagger} \frac{P}{E - H} V .$$  The sum is over all horizon states with which the gravitinos interact.  As a massive particle, the gravitino can only propagate on the null horizon for a proper time of order $m_{3/2}^{-1}$ .  The conventional random walk formula for the propagator shows that in this time it covers an area $\frac{1}{m_{3/2} M}$, where $M$ is a UV cutoff on proper time.  Using Witten's idea\cite{wittenstrong} that the real cutoff is given by the higher dimensional Planck scale, which is the unification scale $M_U \sim 2 \times 10^{16}$ GeV , and the fact that the entropy is the area in four dimensional Planck units we get an estimate of this graph as
$$m_{3/2} \sim R_{dS}^{p} e^{- 2 m_{3/2} R_{dS}} e^{\frac{c M_P^2}{m_{3/2} M_U}} .$$  The first factor represents the power law correction to the exponential estimate and $c \sim 1$.  We've assumed that $V$ has order $1$ (to exponential accuracy) matrix elements to states that constitute a finite fraction of the entropy in the region explored by the gravitino.  We've also used the fact that the horizon states are almost degenerate so the energy denominators don't have much of an effect.  

Now we notice that if $m_{3/2} $ goes to zero like a power of $R_{dS}$ then our formula vanishes exponentially fast if that power is $ p >  -  \frac{1}{2}$, and blows up exponentially fast if $ p < - \frac{1}{2} $.  The only self consistent scaling law for the gravitino mass is 
$$m_{3/2} = a (\frac{m_P}{M_U})^{1/2} \Lambda^{1/4} \sim 10^{-2} {\rm eV} .$$
Here $M_P = \sqrt{8\pi} m_P$ and $a$ is a constant that is nominally of order $1$.  

We do not yet have a derivation of this formula directly from the HST formalism.
It leads to a scale of SUSY breaking in the particle spectrum characterized by an $F$ term of order $20-30 \ ({\rm TeV})^2$, and when combined with the constraints of coupling unification and the absence of new particles, this low scale forces us into a highly constrained set of models for Terascale physics.
I'll outline this in the appendix.

The reader should notice that the role of the c.c. vs. the scale of SUSY breaking is completely reversed in comparison to the two other classes of models studied in this review.  Those models were based on the traditional field theory idea that the c.c. is a low energy effective field theory parameter, to be calculated in some model in which SUSY breaking is fundamental.  Here the c.c. is a fundamental parameter, and its value determines the scale of SUSY breaking, according to the above formula.  One is led to ask, ``then what determines the c.c. ?".

\subsubsection{Cosmology and the Cosmological Constant}

The answer to that question is : cosmology plus anthropic selection.  My own attitude towards the anthropic principle is more or less summarized by Andy Albrecht's motto, "The physicist with the fewest anthropically determined parameters in his/her model, wins".  We know that some things, like why we live on Earth rather than Mars, have only anthropic answers.  We'd like a model of the most basic features of the universe that we observe, which is robust, and doesn't depend too much on the details of biology.  Biologists don't have a very good understanding of how our own type of life arose (presumably) from non-living atoms and molecules.   We have even less chance than they of figuring out whether or not life is possible if the standard model gauge group is replaced by another similar group.   We can really only test these questions for small deviations from the standard models of particle physics and cosmology, or for deviations so large that they lead to some kind of unambiguous disaster.  A particularly good strategy is to produce models in which the alternatives to our own universe are such that there is no period (or too short a period), where localized entropy production due to gravitational collapse, is going on\cite{weinbergbousso}.  It seems reasonable to assume that some such era is necessary to the existence of any kind of complex information processing structure.

Any model, which hopes to utilize anthropic reasoning for some quantities, must contain a mechanism that produces many alternate classical universes, as the result of the action of the same quantum Hamiltonian.  It's also desirable that not too much of the physics in the universe fluctuates from universe to universe in the model.   The couplings we observe are simply incompatible with the hypothesis that they're random variables constrained only by the anthropic principle\cite{bdg}.

The simplest model of cosmology in HST starts from a non-singular Big Bang.    
The trajectories are parametrized by the topology of a cubic lattice with points ${\bf x}$.  The time dependent $H_{in} (t, {\bf x}) $ is independent of ${\bf x}$.  It turns out that in this simple cosmological model the choice of $H_{out}$ is arbitrary, as long as it is ${\bf x}$ independent.  This is a consequence of the fact that the universe described by this model has no local observables.  Recall that in the HST model of Minkowski space, $H_{out}$ for one trajectory was constrained by the local physics described by $H_{in}$ of distant trajectories.  

$t$ is a positive integer (in Planck units) and $H_{in} (t)$ is built from a $t \times (t + 1)$ matrix $\psi_i^A$ and its conjugate.   The full operator algebra is generated by an $N \times (N + 1)$ matrix, with $N \gg 1$.  The variables satisfy
$$[\psi_i^A (k) , \psi^{j\ \dagger}_B (l) ] _+ = \delta_i^j \delta^A_B \delta_{kl}.$$  $H_{in} (t)$ is a function of the bilinears $\psi \psi^{\dagger}$ which is $SU(2)$ invariant.  Apart from this, and the large $t$ behavior prescribed below, we choose $H_{in} (t)$ randomly
at each time $t$.   Of course, when $t = N$ the growth of the operator algebra from which $H_{in} (t)$ is built, stops, and we take the Hamiltonian to be time independent from that time forward.  The effect of this random choice of Hamiltonian is to maximize the entropy at each time, within the Hilbert space ${\cal H} (t) $ generated by the $t \times (t + 1) $ fermions. That is to say, independently of the choice of initial state, the state in this Hilbert space is random.  

The overlap condition is that the overlap Hilbert space satisfies
$${\cal O} (t, {\bf x,y}) = {\cal H} (t - d[{\bf x,y}] ) ,$$ where $d[{\bf x,y}]$ is the minimum number of lattice steps between the two points.  The locus of all points a fixed number of steps from a given point is a tilted cube, and this equation says that, in the space-time metric defined by the causal structure and conformal factor implicit in the QM of HST, this tilted cube is a sphere of radius $t$.

For large time, we insist that the Hamiltonian approach that of a $1 +1 $ dimensional CFT, with central charge of order $t^2$, UV cutoff of order $1/t$, on an interval of order $L t$.  Since the state is random we have 
$$ \langle H \rangle = L t ,$$ while the entropy is
$$L t^2 $$.   $L$ is the number of values taken by the indices $k,l$ in the anti-commutation relations.  It measures the size of the compact dimensions and must be $\gg 1$ for the field theory approximation to make sense.
In the emergent geometry, we have an energy and entropy density 
$$\rho = \frac{L}{t^2} , $$ $$ \sigma = \frac{L}{t} = \sqrt{L\rho} .$$   If $N = \infty$, these are the equations satisfied by a flat FRW model with equation of state $p = \rho$.  This maximally stiff fluid was postulated by Fischler and Susskind\cite{fs} as the equation of state of the very early universe.  

For finite $N$, the asymptotics of the model obviously has the causal structure of de Sitter space.   If we do a rescaling of the $1 + 1$ field theory interval such that the UV cutoff is reduced to $\frac{1}{t^2}$, then the Hamiltonian has the form
$$ H_N = \frac{1}{N^2} {\rm tr}\ P(\psi \psi^{\dagger}), $$ that we proposed for empty de Sitter space in the previous section.  Thus, this model has a coarse grained geometry which is a flat FRW with
   $$ a(t) = \sinh^{1/3} (b t / N),$$ with $b$ a constant of order $1$.   The energy density has two components, with equations of state $\rho_{\pm} = \pm p_{\pm}$.

In this model, the c.c. is determined by $N$, which is an arbitrary input, except for the inequality $N \gg 1$.  As a first step towards a model amenable to an anthropic choice of $N$, we take the previous model at every lattice point belonging to a tilted cube of size $N$ steps, and allow $N$ to go to infinity at every point outside that cube.  We re-define overlaps such that there is no overlap between interior points of the tilted cube and exterior points, and the obvious restriction that the overlap between any exterior point and the boundary is never larger than the finite dimensional boundary Hilbert space.   Other overlap conditions remain unchanged, and the consistency conditions are satisfied.

The obvious geometrical interpretation of this is that the boundary is a marginally trapped surface embedded in the $p=\rho$ geometry, and the interior is a de Sitter space with the same horizon area.   The transition between the two takes place over a Planck distance, and there are solutions of the Israel junction conditions, with a sensible boundary stress tensor, for such a geometry.  

There are obviously solutions (in the Israel sense, at least) to Einstein's equations describing multiple black holes, of varying sizes, initial positions and velocities embedded in the $p=\rho$ FRW, as long as the initial separations are much larger than any of the Schwarzschild radii.  Thus, although we do not yet have a quantum model corresponding to such solutions, we anticipate a framework where the c.c. varies over this multiverse and we can apply anthropic reasoning\footnote{The eventual black hole collisions that are implicit in this setup also provide one of the virtually infinite number of ways to solve the so called Boltzmann Brain problem, in a model with a stable dS space, without affecting any observable prediction of the model.}.  

In the framework of HST, the connection between SUSY breaking and the c.c.  makes anthropic reasoning look quite different from traditional arguments.
There is no time in this review to go through the detailed arguments that lead to a particular class of models, the Pyramid Schemes\cite{pyramid}, which implement this connection.  Instead I'll try to indicate the general thrust of the arguments. One starts from a model with zero c.c., super-Poincare invariance, and a discrete complex R symmetry.  The low energy physics of this model involves a gauge group $G$ and a reducible representation for chiral multiplets.  There are no moduli. There are no known models of this type in string theory, and experience with string theory shows that the imposition of a discrete R symmetry is a significant constraint.  Thus, it is plausible that there are not many models of this type. Questions about which gauge groups and representations are allowed cannot be answered without a much more detailed understanding of the dynamics of HST in Minkowski space.  We will approach them phenomenologically, insisting on the standard model gauge group and representation content.  Anthropic arguments will be limited to questions of whether the scales of these interactions can be changed.
Note that the UV values of gauge couplings and quark and lepton couplings to the Higgs will have finite limits as $N \rightarrow \infty$ and are thus insensitive to the value of the c.c. .

Interactions with the horizon DOF in dS space, which decouple in the limit of vanishing c.c., breaks the R symmetry and gives rise to a gravitino mass $m_{3/2} \sim (\frac{m_U}{m_P})^{1/2}  N^{ - \frac{1}{2}} M_P $.   In effective field theory, this is implemented by adding R violating terms to the Lagrangian, which lead to spontaneous SUSY breaking\footnote{SUSY is a gauge symmetry in SUGRA, so all breaking of it is spontaneous.  The real significance of the nomenclature is the fact that the scale of SUSY breaking can be dialed to zero as the c.c. vanishes, so that the super-partners of the Goldstino have to appear in the low energy Lagrangian.}.  
The only way I know how to do this is to use the O' Raifeartaigh mechanism for SUSY breaking.  

The gauge group $G$ must include the standard model. The O' Raifeartaigh fields have to include standard model singlets, and might include the Higgs fields\footnote{One cannot include lepton superfields at the renormalizable level.  As a consequence of the R symmetry, the model is lepton flavor conserving at the renormalizable level when $\Lambda = 0$, and the diagrams that violate R symmetry involve gravitino exchange with the horizon and conserve lepton flavor.}.
Other candidates among the standard model superfields are excluded by symmetries (see previous footnote).  If $G = SU(1,2,3)$, SUSY breaking effects 
in the standard model come only through coupling to the Higgs fields.  The light quark/lepton/gauge sector is approximately supersymmetric\footnote{In fact, at the renormalizable level in this model, SUSY breaking only affects a singlet chiral field, which decouples.}, with disastrous consequences for atoms, and through them for stars.  The absence of humans is a minor consequence of these cosmological disasters.   To avoid this we must introduce another factor in the gauge group, $G_P$ and chiral superfields $T_i, \tilde{T}_i$ that transform under both $G_P$ and the standard model\footnote{In the original papers\cite{pyramid}, these fields were introduced for phenomenological reasons.}.  The mass terms for the $T$ fields violate the discrete $R$ symmetry and scale like $N^{- 1/4} M_P$.

In the Pyramid scheme, $G_P = SU(3)$ and the model is assumed to be near an approximate fixed line with fairly strong gauge coupling as it flows from the UV\footnote{The $T_i, \tilde{T}_i$ are in the $(3,\bar{3}_i) \oplus (\bar{3}, 3_i)$ of the product of $G_P$ and Glashow's trinification\cite{trinification} group.}.  In the $N \rightarrow\infty$ limit, this behavior persists into the deep IR, though the model eventually becomes IR free.  When $N$ is finite the gauge bosons of $G_P$ become strongly coupled and confine at a scale $\Lambda_3$ which is of order
the mass of the lightest $T_i$ field.  The Hubble scale $N$ thus affects the scale of SUSY breaking and electro-weak symmetry breaking, and modifies the RG equations of all the standard model couplings.  Since dark matter is identified with a baryon made from the $T_i$ fields, varying $N$ changes the properties of dark matter as well.  It is extremely plausible that, within this class of models, $N$ is fixed by the requirement that atoms and stars behave more or less as they do in the real world.  

Our simple models of cosmology are unsatisfactory because they do not contain any localized excitations.  Time evolution maximizes the entropy at all times, in the sense that the density matrix averaged over a few Planck times is highly impure.
Some readers may be surprised at this description of empty dS space, but we have long known that empty dS space had large entropy\cite{gibhawk}.  Furthermore, introducing a classically stable localized excitation, in the form of a black hole, {\it reduces} the entropy.  For black hole masses much smaller than the Nariai maximum\cite{nariai}, this gives precisely the thermal suppression expected from the dS temperature.  Our models of particle physics in Minkowski and dS space were built to mimic this behavior.   Particle states in a large causal diamond are highly constrained, low entropy states, while empty space has maximal entropy.
Our conventional count of entropy leaves out the entropy of the boundary of our causal diamond,and the fact that localizable excitations constrain some of the boundary DOF to vanish.

The problem then is to modify our model of evolution from $p = \rho$ to $p = - \rho$
cosmology in such a way that local excitations of empty dS space are generated at an intermediate time.   The clue to how to do this comes from the conventional theory of inflation, though our eventual model differs from a field theoretic inflation model in profound ways.  

Consider the model we've described in detail, with the number $N$ now thought of as the inflationary Hubble radius in Planck units.  Accordingly, we will rename it $n$ and reserve $N$ for the actual Hubble radius to which our universe asymptotes.
Choose a particular point on the lattice of trajectories and call it the origin.  Trajectories that are more than $n$ steps away on the lattice have no overlap at all with the trajectory at the origin.  Thus, we can make a coarse grained lattice of points a distance $n$ away from each other, such that the DOF associated with those points are completely unconstrained and independent.  This is what we expect for disjoint horizon volumes in dS space.  In this model thought, they are simply copies of the same information that's available in a single horizon volume, which is in accord with the Covariant Entropy Principle and the Banks-Fischler conjecture that dS space has a finite number of states.  The idea of the holographic inflation (HI) model is to incorporate $e^{3N_e}$ of these copies into the Hamiltonian of a single trajectory, and allow them to interact gradually.  

We have not yet confirmed that we can construct a fully consistent HST model, with overlap conditions for trajectories separated by a Planck distance, which has a single trajectory Hamiltonian equal to that of the HI model.   In the HI model we concentrate on $e^{3Ne} $ trajectories, ${\bf x_i}$ separated by $n$ lattice steps, and write a Hamiltonian acting on the tensor product of all their Hilbert spaces.  Initially the Hamiltonian is  $$ H (t_{be}) = \sum_{k=1}^{e^{3N_e}}  H_n ({\bf x_i}) ,$$ where the individual Hamiltonians are identical in form but act on different tensor factors.   The time $t_{be}$ is larger than $n$ .   We call it {\it the beginning of the end of inflation}.  

For $t > t_{be}$ we begin to couple these systems together.    We array the $e^{3N_e} n (n + 1) $ DOF on a ``fuzzy 3-sphere".  The details of that construction may be found in \cite{holoinflation}, but roughly speaking it assigns $n(n+1)$ DOF to a grid on the 3 sphere with spacing $e^{- N_e}$.  In the limit of large $N_e$ we can construct generators which converge to the algebra of $SO(1,4)$ out of these variables, which behave like commuting ({\it i.e.} the bilinears commute) localized fields on the 3-sphere.  The $SO(3)$ subgroup of $SO(1,4)$ is an exact symmetry at all times, and the other generators emerge in the large $N_e$ limit.  The Hamiltonian $H(t)$ varies slowly from $H (t_{be})$ to the approximate $SO(1,4)$ generator $J_{04}$, which commutes with $SO(3)$.  As time goes on, the couplings between the point we have called the origin and points close to it on the 3-sphere, are turned on, eventually covering the whole sphere at some time $t_{e}$, the {\it end of inflation.}

As we've said, we have not implemented the full consistency conditions of HST, but the approximate $SO(4)$ symmetry guarantees that we could have chosen any trajectory on the three sphere as the origin.  That is, since our model is something like a lattice field theory, we can implement the consistency conditions as they would be implemented in such a field theory.  

By construction, the physical radius of the horizon at the end of inflation, satisfies $E^2 \sim e^{3N_e} n^2 .$.   We can use this relation to constrain the space-time picture associated with the model.  It is a flat FRW model, which extends the original $$ a(t) = \sinh^{1/3} (ct) ,$$ to a slow roll expansion, in which the horizon gradually expands.  The choice of time dependence of the quantum Hamiltonian $H(t)$ determines the form of $a(t)$ .   Our model is that the coupling together of DOF expands out from the pole of the 3-sphere (which we've called the origin) at a certain rate, and that at any given time, those DOF not yet coupled to the origin, continue to evolve with their independent copies of the Hamiltonian $H_n$.  Eventually we reach the Hamiltonian $$H( t_{ee}) = J_{04} ,$$ and all the DOF are coupled together.  

The HST consistency conditions before $t_{be}$ guarantee that each horizon volume is in the same initial state at this time. The mismatch between the time dependence of $H(t)$ and the unperturbed evolution of as yet uncoupled horizon volumes under $H_n$ introduces fluctuations in the state.   Since individual points on the 3-sphere correspond to individual systems of entropy $\sim n^2$, the fluctuations will be localized.  If $n$ is large, the $k$ point connected correlation function of the fluctuations will be $o(n^{-k})$.  If $N_e$ is large, the system will have an approximate representation of $SO(1,4)$ symmetry, and the time averaged density matrix
$$ \bar{\rho} = \frac{1}{T} \int\ dt\ e^{- i J_{04} t} \rho e^{i J_{04} t} ,$$ will be approximately invariant.   The $k$ point functions are
$${\rm Tr}\ [ \bar{\rho} S(x_1 ) \ldots T(x_j) \ldots ] .$$ Here $S$ and $T$ are commuting operators on the 3-sphere (they commute because they are even functions of independent fermionic DOF), which represent the scalar and tensor fluctuations in cosmological perturbation theory.  We are here taking the Jacobsonian point of view that cosmological perturbation theory is the hydrodynamic theory of fluctuations in the underlying quantum model.
The operators $S$ and $T$ are assumed to transform as unitary representations of $SO(1,4)$.  The tensor representation is unique, but the scalar can belong to any one of the complementary series of unitary representations\footnote{In single field slow roll QFT models of inflation, $S$ transforms as the singular limit point of the complementary series.} .   It's possible that a better understanding of the microscopic theory will fix the scalar representation more precisely.
Of course, $SO(1,4)$ covariance cannot hold for arbitrary $k$ point functions, because our system has only a finite dimensional Hilbert space, but the dimension is ${\rm exp} (e^{N_e}) $, so it is reasonable to assume that this cutoff does not have a significant effect on low point functions, which are the only predictions we can compare to data.

In \cite{holoinflation2} we showed the approximate $SO(1,4)$ invariance and the general rules of cosmological perturbation theory, including Maldacena's\cite{malda} squeezed limit theorem, was enough to reproduce the data.  This general framework predicts a nearly scale invariant scalar two point function, and a small, equilateral scalar three point function, which is probably below the limits of near future observations.  The tensor to scalar ratio $r$ is predicted to be small, by a factor $\frac{\dot{h}}{h^2}$, where $h = \frac{\dot{a}}{a}$.  The HST and single field slow roll predictions differ both because of the freedom of choice of $SO(1,4)$ representation in HST, and because the slow roll quantum fluctuations are evaluated in the adiabatic QFT vacuum, which is not an $SO(1,4)$ invariant density matrix.   However, this can be masked by taking a different choice of $a(t)$ in the two models.

At the level of the tensor two point function there is an unmistakable difference between the two models.  HST predicts vanishing tensor tilt.  Unfortunately, slow roll models predict a tilt of $r/8$, and $r$ is bounded experimentally by $\sim 0.1$.
The single field slow roll consistency condition between scalar and tensor amplitudes and tilts is not expected to be satisfied in HST.  There are even larger discrepancies between the predictions of the two models for tensor 3 point functions, but these will not be measured in the foreseeable future.

The bottom line is that the holographic inflation model is as good a fit to current data as any inflationary model, and that it avoids all of the Trans-planckian puzzles of field theoretic inflation, and puts the theory of inflation into a finite and consistent quantum theory of gravity.  What the theory is missing is an analog of the mechanism of reheating.  In previous sections I've outlined the HST models of Minkowski and dS space and showed how particle physics arises from the spinor pixel variables of HST.  Presumably, the local excitations that arise in the holographic inflation model, transmute into the particles of our model of dS space, since they are built out of the same variables, but the details of this process, which bear on questions like baryogenesis, the origin of the primordial dark matter density, and a number of other cosmological parameters, are still obscure.

As it stands, the model has two parameters, $n$, the Hubble radius during inflation, and $N$, the Hubble radius to which we appear to asymptote.  They are constrained, in the model, only by the {\it a priori} inequalities $1 \ll n \ll N$. Above, I've argued that, because of the connection between $N$ and SUSY breaking, it is likely that $N$ is strongly constrained by anthropic (really atomthropic) constraints on particle physics.   It also seems likely that the dark matter density $\rho_D$ at the beginning of the matter dominated era will be determined by similar considerations.   In that case, Weinberg's\cite{weinbound} galactothropic constraint on cosmological parameters may be written (including the {\it a priori} constraints)
$$1 \ll \frac{n}{x} <  N^{ \frac{2}{3}} (\frac{729 \rho_D}{500 M_P^4})^{1/3} \ll N.$$  $x$ is a constant relating the primordial  energy density fluctuations to $n$
$$ \frac{\delta\rho}{\rho} = \frac{x}{n} .$$ It is the inverse of a slow roll factor and the CMB and large scale structure data indicates that $ x \sim 10$ and $\frac{n}{x} = 10^5$.   Thought of as a bound on $n$, with $N$ and $\rho_D$ fixed, the RHS is remarkably close to this observed value.

To summarize: HST cosmology provides a framework in broad agreement with all cosmological data, if we surmise that post-inflationary reheating works in a conventional manner.  It is completely quantum mechanical and non-singular, and accounts for the low entropy initial state of the universe by invoking a period of inflation to create localized fluctuations.  The idea that localized fluctuations are low entropy follows from properties of dS space. Homogeneity, isotropy and flatness follow from generic initial conditions and need no further inflationary explanation.
The inflationary and teleological Hubble scales are both environmentally selected parameters.   The c.c. is determined by atom/star-thropic constraints, while Weinberg's galactothropic constraint puts an upper bound on the inflationary Hubble radius, which is constrained to be large.
The two most unusual features of this cosmology are the connection between localized excitations, low entropy and inflation, and the way in which the connection between the c.c. and SUSY breaking puts strong constraints on the c.c. .

To conclude this section, I'd like to comment on the question of how the HST models compare to the String Landscape with respect to the question of anthropic determination of parameters.  In HST, the c.c. and $\frac{\delta\rho}{\rho}$ vary over the Landscape.   Everything else is determined by the super-Poincare invariant limiting model, to which the $dS_N$ universe converges as $N \rightarrow\infty$.  That model is fixed by the choice of anti-commutation relations for the fundamental variables $\psi_i^A (P)$.  Each discrete choice of compact fermionically generated super-algebra, which admits a super-Poincare sub-algebra in the large $N$ limit, is a potential candidate, but experience from string theory suggests that the existence of a super-Poincare invariant S-matrix will put further constraints on the model.
There is no sense, in HST in which these choices are scanned by looking at different meta-stable states of the system.  This agrees with the CDL analysis, in which there is no tunneling between super-Poincare invariant states, as well as the black hole based arguments of \cite{tbvac}.  In HST only two parameters are scanned in the Landscape.

If it turns out that there are many different mathematically consistent HST models, all of which are consistent with anthropic constraints, then we will be in the situation that theoretical physics 
has been in throughout its history.  The mathematical tools with which we analyze the universe produce many imaginary worlds, which seem no less consistent than the one that actually describes the universe.

\section{L' Envoi}

I have reviewed three attempts to explain the marvelously small value of the c.c., and their connection to SUSY.
Of the three, the one which leaves me least impressed is the most popular one, The String Landscape.  The Landscape starts from a very plausible construction of a large class of supersymmetric $AdS_4$ models of quantum gravity, which are based on a small modification of Poincare invariant F-theory solutions of Type IIb SUGRA.  Cancellations of different terms contributing to the constant $W_0$ in the super-potential, allow a small subset of these models to have tiny negative values of the c.c. . The AdS/CFT correspondence tells us that the rigorous definition of these models is a large collection of independent super-conformal field theories.

The theoretically implausible part of the landscape story is the process of ``uplifting", which is simultaneously averred to be a small perturbation of the previous model, but also a model in which all of the different CFTs have suddenly become different meta-stable states in the same quantum model, whose relationship is governed by semi-classical tunneling amplitudes.  To me, these two claims seem incompatible.   It should also be emphasized that none of this is really String Theory, in the sense that there is some sort of (order by order) finite world sheet perturbation expansion of well defined observables.   Instead it is string inspired effective field theory, or string inspired effective field theory with singular D-brane sources.  The only attempt at a mathematical definition of observables for the uplifted model is the FRW/CFT correspondence.  That theory is at best conjectural, and the relation of its mathematical observables to measurements in our own universe is completely obscure.

In addition to problems of principle, the String Landscape has a host of phenomenological problems.  We've covered these in the text.   The question of the relation between the c.c. and SUSY breaking has received a variety of answers in the literature on the Landscape.  I believe a number of results point to the answer that a combination of anthropic and meta-stability arguments drive the most probable allowed universe to have low energy SUSY.  This exacerbates the phenomenological problems.  

SLED theories are much less ambitious than the String Landscape.  They attempt to solve the c.c. problem in the context of low energy six dimensional effective SUGRA.
What they do succeed to do is to relate both the gauge hierarchy problem and the c.c. problem to a single fact:  why are the extra dimensions so large?  As far as I understand it, 
this question can only be answered in a UV completion of the six dimensional theory.   I don't think there can be any purely dynamical answer to it, since the only parameter in the theory is the six dimensional Planck scale.  This again points to a Landscape of possibilities, and it is hard to address the question without establishing the nature of that Landscape.  For example, if it were to turn out that the UV theory had an additional hidden parameter which determined the size of the extra dimensions in Planck units, one would have to determine whether different values of that parameter corresponded to different models, or different states in the same model.  In the latter case, we'd have to determine the mathematical prediction for the probabilities of each value.
In either case we'd have to answer phenomenological questions about why certain operators on the four dimensional brane are missing from the world around us, but in the case where the models were distinct we could invoke symmetry arguments to explain most of these miracles.

HST models of cosmology reject the notion that the c.c. is a calculable parameter affected by low energy loops of a QUEFT.  Instead, following Jacobson, it views space-time geometry (and, in the string theory/higher dimensional SUGRA context, this includes all the gauge and matter fields of the standard model) as an arena in which physics unfolds.  The fundamental formalism has quantum avatars of time-like trajectories and causal diamonds built into its structure, as well as an ``equivalence principle" defining consistent shared information, which knits the separate trajectories into a space-time.  Classical GR, except for the c.c., is an expression of the local thermodynamics ($=$ hydrodynamics) of this quantum system, and the c.c. is an asymptotic boundary condition relating the asymptotics of proper time to the asymptotics of area ($\propto$ entropy).

In the context of exactly symmetric spaces, the c.c. is thus a quantized input parameter.  In a broader cosmological context, I've described a model which produces a $p = \rho$ FRW, with an arbitrary number of bubbles, whose exterior looks like that of a black hole of equal entropy to the interior asymptotically dS space.  The interior of each bubble describes a cosmology that begins with a $p = \rho$ era, passes through a period of slow roll inflation, with Hubble radius $n \gg 1$ in Planck units, and asymptotes to dS space with Hubble radius $N \gg n $.  The inflationary era explains the emergence of local physics, which is simultaneously the explanation of the low entropy initial state of the universe.  

A large collection of bubbles allows for environmental selection of those which allow for the formation of complex structures and energy transport\footnote{A wide selection of initial conditions for the relative positions and velocities of the bubbles, also solves the Boltzmann Brain problem, which could be solve in an infinite number of different ways, none of which change the predictions of the model for observable physics.}.  The question of whether other parameters of the low energy QUEFT are ``scanned" is moot, because these are encoded in the commutation relations of the fundamental values, and each choice corresponds to a different model, rather than a different quantum state of the same model.

However, it is certainly true that the number of possible models is much more constrained than the number of AdS models that form the basis of the string landscape.  
At the level of effective field theory each of these models corresponds to a solution of $$ D_i W = W = 0 ,$$ which is an over constrained system of equations.  One must also impost a condition on the discrete R charges of chiral fields,
$$N_2 \geq N_0 ,$$ where $N_q$ is the number of fields with R charge $q$.  This guarantees that there is no continuous moduli space, another feature of the $\Lambda = 0$ limit of HST models.  

In the equations of specific SUGRA compactifications, the condition of discrete R symmetry has many fewer solutions than the equations $D_i W = 0$\cite{dz}, essentially because most of the fluxes that give rise to the huge multiplicity of solutions, also break R symmetry.  Furthermore, all known solutions of this type have moduli, at least the parameter that justifies the classical approximation.  To leading order in that parameter, one cannot find an R symmetric solution, which freezes the parameter, even with the tuning that justifies the expansion.   It seems reasonable to conclude that the number of possible HST models, for each value of $n$ and $N$, is quite small.

\vskip.3in
\begin{center}
{\bf Acknowledgments }
\end{center}
 I would like to thank M.Douglas, E.Silverstein, S.Kachru, L.Susskind, S.Shenker, C.Burgess, R. Bousso, and especially M. Dine, for numerous conversations about the topics covered in this review. This work was supported in part by the Department of Energy.

\end{document}